\documentclass[a4paper,fleqn,usenatbib]{mnras}

\usepackage[T1]{fontenc}
\usepackage{ae,aecompl}
\usepackage{lipsum}
\usepackage{mathtools}
\usepackage{cuted}
\usepackage{float}

\usepackage[authoryear]{natbib}

\usepackage{graphicx}	% Including figure files
\usepackage{amsmath}	% Advanced maths commands
\usepackage{amssymb}	% Extra maths symbols
\usepackage{breqn}

\usepackage{multirow}
\usepackage{booktabs}
\usepackage{amsmath}

\title[The early evolution of Type I GCs]{SNe and their impact during the early evolution of Type I Globular Clusters}

\author[Jim\'enez et al.]{
Santiago Jim\'enez,\thanks{E-mail: sjimenez@inaoep.mx}
Guillermo Tenorio-Tagle,
Sergiy Silich
\\
Instituto Nacional de Astrof\'isica, \'Optica y Electr\'onica, AP 51, 72000 Puebla, M\'exico\\
}

\date{Accepted XXX. Received YYY; in original form ZZZ}

\pubyear{2020}

\begin{document}
\label{firstpage}
\pagerange{\pageref{firstpage}--\pageref{lastpage}}
\maketitle

\begin{abstract}
The iron composition of globular clusters (GCs) is homogeneous in all but a few massive clusters, despite the presence of multiple stellar populations. Hence, most if not all the supernovae (SN) ejecta was not used to form stars. Here by means of semi-analytic and numerical studies we address this issue considering both stellar winds and supernovae feedback during the early evolution of proto-globular clusters. We calculate the ability of stellar winds to form a global wind that removes the gas left over from star formation. The innermost radius from which such a global wind can be formed, the superwind radius $R_{SW}$, is a function of the cloud parameters and the star formation efficiency. In the case of complete gas expulsion ($R_{\textrm{SW}}=0$), the SN ejecta merge with shock-heated winds and exit the cluster. On the other hand, when $R_{\textrm{SW}}>0$, supernova remnants (SNRs) become pressure-confined if evolving within a critical radius $R_{\textrm{blow}}$, and mix their products with the residual gas. However, outside of this central zone the SNRs experience blowout. In such cases, the thermalized ejecta escapes the cluster, making the SN products unavailable for the formation of new stars. We estimated the metallicity enhancement ($\Delta\textrm{[Fe/H]}$) of the leftover gas and discuss the conditions required to produce secondary stellar populations with $\Delta\textrm{[Fe/H]}$ in the range observed in the majority of GCs.

\end{abstract}

\begin{keywords}
globular clusters: general - galaxies: star clusters: general - ISM: supernova remnants. 
\end{keywords}

%%%%%%%%%%%%%%%%%%%%%%%%%%%%%%%%%%%%%%%%%%%%%%%%%%

%%%%%%%%%%%%%%%%% BODY OF PAPER %%%%%%%%%%%%%%%%%%

\section{Introduction}
Multiple stellar populations (MPs) have been discovered to be common within ancient galactic globular clusters (GCs) through extensive spectroscopic and photometric studies over the last decades  (see \citealt{1999Natur.402...55L,2004ApJ...605L.125B, 2012A&A...544A..12G}, \citealt{2015A&A...578A.116C}, \citealt{2015AJ....149...91P}, \citealt{2015MNRAS.447..927M}, \citealt{2015ApJ...808...51M}, \citealt{2019MNRAS.487.3815M} and references therein). Recently the MPs phenomenon was also confirmed for GCs in nearby local group galaxies such as the Large (LMC) and Small (SMC) Magellanic Clouds \citep[e.g.][]{2007AJ....134.1813M, 2009ApJ...695L.134M, 2016ApJ...829...77D,2017MNRAS.465.4159N, 2019MNRAS.486.5581G, 2019MNRAS.487.5324M}, and the Fornax and Sagittarius dwarf galaxies \citep[e.g.][]{2014ApJ...797...15L,  2019MNRAS.490L..67S,2021arXiv210207785F}. The multiple populations (MPs) clusters are one of the most unexpected findings in stellar astrophysics, which has converted the formation and evolution of GCs into one of the most challenging research topics.\\ 
\indent From the observational point of view there are two types of GCs (see \citealt{2019MNRAS.487.3815M,2019A&ARv..27....8G}):  Both Types present two main groups of stars. In one of them, the stars share the same light-element content as the field stars, so it is thought they formed from their proto-cluster primordial cloud and are called the first (1G) stellar generation. The second group of stars (called 2G) present chemical anomalies, the most recurrent ones being the clear anti-correlations Na-O and N-C. Other observed chemical signatures of 2G stars are, among others, the  Al-Mg anti-correlation found in massive metal-poor clusters \citep[e.g.][]{2017A&A...601A.112P, 2019AJ....158...14N}, and higher He abundances. However, in Type I clusters, which comprises 83 per cent of the objects, both generations are homogeneous regarding  the Fe group elements, whereas Type II GCs are more complex, with a fraction of 2G stars enhanced in Fe and with lower Li abundances \citep[e.g.][]{2008ApJ...672L..29Y, 2009ApJ...705.1481D, 2015AJ....150...63J,2017ApJ...836..168J, 2015MNRAS.450..815M,2019MNRAS.487.3815M}. Such detailed features, based on a vast amount of observations, has led to many theoretical issues, some of which seem insurmountable (see \citealt{2018ARA&A..56...83B} and references therein). For instance, it is still unclear how to generate the conditions to form new stars in a place already occupied by a major burst of star formation, or what masses and spatial distributions of the parental gas clouds are necessary for the development of such chemical peculiarities. \\
\indent It has been suggested that in order to explain the light elements anti-correlations, the gas to be used for a second generation has to be polluted with H-burning products produced at high temperatures within the interior of 1G stars \citep{2002A&A...395...69D}. There are several candidates that potentially can pollute the intra-cluster cloud with H-burning products: AGB stars (\citealt{2004ApJ...611..871D}, \citealt{2010MNRAS.407..854D}, \citealt{2016MNRAS.458.2122D}), massive binaries (\citealt{2009A&A...507L...1D}, \citealt{2013MNRAS.436.2398B}, \citealt{2019ApJ...879...58T}), fast rotators (\citealt{2007A&A...464.1029D}, \citealt{2007A&A...475..859D}) or super massive stars (\citealt{2014MNRAS.437L..21D}, \citealt{2018MNRAS.478.2461G}). However,  all of these possibilities struggle to explain all the observational constraints found up to date.

Star clusters go through an initial hidden phase of their evolution. So the first stellar generation evolves buried in the gas left over from  star formation. The length of the hidden phase and whether the leftover gas can mix with chemical enriched material to form a secondary stellar population, depend on the ability of the stellar feedback in pushing the leftover gas out of the cluster (see \citealt{2020SSRv..216...64K}  and references therein). Stellar winds, ionizing radiation and SN explosions inject energy, mass and momentum into the intra-cluster medium, which can increase the gas pressure and lead to gas expulsion through a global wind \citep[e.g.][]{2006MNRAS.373..752G, krause2013superbubble, 2015ApJ...814L..14C, 2015MNRAS.451..987D, 2017MNRAS.470.4453R, 2019MNRAS.489.3269C}. However, currently the efficiency of such gas removal is not clear. Indeed, the high densities and pressures inside massive and compact clouds may inhibit the formation of a central global wind \citep[e.g.][]{Silich2017,Silich2018,Silich2020}, thus suppressing gas removal. Moreover, \cite{2018MNRAS.476.5341F} show that substructure within the cluster can also reduce the efficiency of gas expulsion. 

Therefore, the core-collapse SN feedback (which happens between 3 to 40 Myr) may take place during the hidden evolutionary stage of massive and
compact clusters. Here we center our attention on this issue. We explore the possibility that the blowout of supernova remnants out of the cloud left over from star formation is the key to understand why most GC show the same Fe metallicity in all their stellar generations. \\
\indent Blowout was thoroughly investigated during the 1980s and 90s for the case of galactic superbubbles fed by stellar winds and SNe \citep[e.g.][]{1985ApJ...299...24S, 1986PASJ...38..697T, 1987A&A...179..219T, 1987A&A...182..120T, 1988ARA&A..26..145T, 1989ApJ...337..141M, 1992ApJ...388...93K,silich1992, 2013A&A...557A.140B}. In this case Rayleigh-Taylor (RT) instabilities lead the hot gas (the thermalized SN ejecta and stellar winds) to blowout into the galactic halo of the host galaxy.

Here we consider massive and compact proto-cluster clouds and investigate whether individual supernova remnants (SNRs) resultant from former stellar generations, may also experience blowout from the cloud left over from star formation \citep{tenorio2015supernovae}, inhibiting then its contamination. 

The organization of this paper is as follows: section \ref{clusterModel} describes the star forming cloud model (the way stars and gas are assumed to be distributed). Section \ref{wind_feedback} deals with the stellar winds feedback. Massive compact clusters are shown to retain most of the leftover gas, except in the outermost regions, where a cluster wind may form and remove some of the gas. The radius where such a gas removal is possible, here called the superwind radius $R_{\textrm{SW}}$, is calculated as a function of the cloud parameters. Section \ref{mainEquations} presents the considered physics of SN explosions and their implementation in a 3D thin-shell approximation. Section \ref{typicalBlowout} throughly describes a typical blowout event. Later, section \ref{blowradii_sec} presents the calculation of the blowout radii $R_{\textrm{blow}}$ for a set of clouds with different masses and sizes. The blowout radius is the minimum distance away from the cloud center where a SN explosion would lead to an SNR able to accelerate, fragment and release the ejecta into the surroundings. For SN explosions within this radius, the high density and turbulent pressure lead to pressure confinement and thus to their trapping within the cloud. This allows for a possible Fe contamination of the gas out of which the 2G will form. The metallicity enhancement due to these trapped SNRs is discussed in section \ref{enrichment_section}. Finally, a summary and conclusions of our results are presented in Section \ref{summarySection}.
\section{The star-forming cloud}\label{clusterModel}
We consider the early evolution of proto-cluster clouds, such that 1G stars with mass $M_{\textrm{1G}}=\epsilon_{\textrm{1G}} M_{\textrm{tot}}$ remain embedded into the leftover gas whose  mass is $M_{\textrm{gas}}=\left(1-\epsilon_{\textrm{1G}} \right)M_{\textrm{tot}}$, where $M_{\textrm{tot}}$ is the total mass of the star-forming cloud and $\epsilon_{\textrm{1G}}$ is the 1G star formation efficiency.
Hereafter a Gaussian density distribution is assumed for both the gas left over and stars:
\begin{equation}\label{eq1}
\rho_{g} \left(r \right)=\left\{
	\begin{array}{ll}
		\frac{\left(1-\epsilon_{\textrm{1G}} \right)M_{\textrm{tot}}}{\left( 2 \pi\right)^{3/2}R_{\textrm{c}}^{3}} \exp \left[-\frac{1}{2}\left( \frac{r}{R_{\textrm{c}}}\right)^{2} \right],  & \mbox{if }  r \leq  R_{\textrm{SC}},\\
     \rho_{\textrm{amb}} & \mbox{if } r > R_{\textrm{SC}},
	\end{array}
\right.
\end{equation}
\begin{equation}\label{eq1a}
n_{*} \left(r \right)=\left\{
	\begin{array}{ll}
		\frac{N_{\textrm{1G}}}{\left( 2 \pi\right)^{3/2}R_{c}^{3}} \exp \left[-\frac{1}{2}\left( \frac{r}{Rc}\right)^{2} \right],  & \mbox{if }  r \leq  R_{\textrm{SC}},\\
     0 & \mbox{if } r > R_{\textrm{SC}},
	\end{array}
\right.
\end{equation}
where $R_{c}$ is the core radius, $R_{\textrm{SC}}$ the cloud boundary, $\rho_{\textrm{amb}}=\mu n_{\textrm{amb}} $ is the uniform density of the surrounding medium and $\mu=14/11 m_{H}$ the mean mass per particle in the neutral gas with 10 hydrogen atoms per helium atom. $N_{\textrm{1G}}$ in equation (\ref{eq1a}) is the total number of 1G massive stars 
\citep[e.g.][]{2015ApJ...814L..14C}:
\begin{equation}\label{massive_stars_total}
N_{\textrm{1G}}=10^{4}\frac{M_{\textrm{1G}}}{10^{6} M_{\odot}}.
\end{equation}
This equation assumes that stars form with a Kroupa initial mass function with lower and upper mass cutoffs equal to 0.01 M$_{\odot}$ and 100 M$_{\odot}$, respectively. 

The cloud gas pressure is determined by the equation:
\begin{equation}\label{eq2}
P_{g}\left(r \right) =\left\{
	\begin{array}{ll}
		-\int_{R_{sc}}^{r} \frac{G M\left(r \right)\rho_{g} \left(r \right)}{r^{2}} dr +P_{\textrm{amb}},  & \mbox{if }  r \leq  R_{\textrm{SC}},\\
     P_{\textrm{amb}} & \mbox{if } r > R_{\textrm{SC}},
	\end{array}
\right.
\end{equation}
where $G$ is the gravitational constant, $P_{\textrm{amb}}=k n_{\textrm{amb}}T_{\textrm{amb}}$ is the thermal pressure of the ambient gas, $k$ is the Boltzmann constant and $M\left( r\right)$ is the 
total mass contained within the radius $r$:
\begin{equation}\label{eq3}
M\left(r \right)=\frac{4 \pi}{1-\epsilon_{\textrm{1G}}} \int_{0}^{r}  \rho_{g}\left(r \right)r^{2} dr. 
\end{equation}
It was assumed that $n_{\textrm{amb}}=10^{-3}$ cm$^{-3}$ and $T_{\textrm{amb}}=10^{4}$ K in all the calculations. Next, the one-dimensional velocity dispersion $\sigma$ is calculated with the equation $P_{g}=\rho_{g}\sigma^{2}$.\\  
\indent Note that equation (\ref{eq2}) can be integrated analytically to find an expression for the central pressure $P_0$ in a Gaussian cloud (see Appendix \ref{Ap2}):
\begin{equation}\label{turbPfinalINI}
P_{0} = 0.17 \frac{G M_{\textrm{gas}}^2}{\left( 2 \pi \right)^{3/2}\left(1-\epsilon_{\textrm{1G}} \right)R_{c}^{4}}.
\end{equation}

\begin{figure}
	\includegraphics[width=1.\columnwidth]{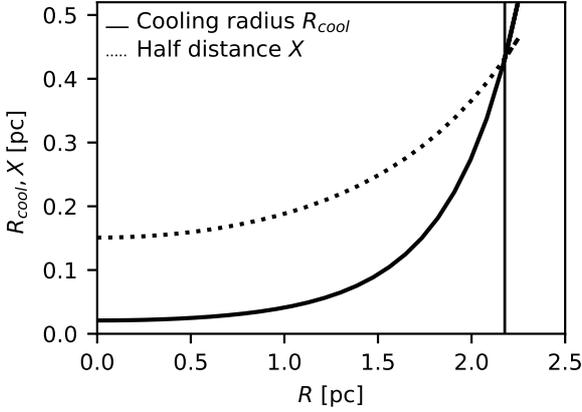}
        \caption{The cooling radius (solid line) and the half distance $X$ (dotted line) as a function of the position within the gas cloud. The vertical thin solid line shows the distance where $R_{\textrm{cool}}/X=1$. The cloud parameters for this case are $M_{\textrm{tot}}=7.14 \times 10 ^{5}$ M$_{\odot}$, $\epsilon_{\textrm{1G}}=0.1$, $R_{c}=0.87$ pc and metallicity $Z=10^{-2}$ Z$_{\odot}$. }
    \label{fig2_winds}
\end{figure}

\section{Stellar winds feedback}\label{wind_feedback}
\subsection{The superwind radius}
Before SN explosions, the mechanical feedback in the cluster is dominated by stellar winds from massive stars. A global cluster wind that expels the residual gas from the cluster is formed if individual wind-driven bubbles collide (e.g. \citealt{2003MNRAS.339..280S}, \citealt{2014MNRAS.442.2701R}, \citealt{2018MNRAS.478.2794N}  and references therein). However, \cite{Silich2017,Silich2018,Silich2020} recently pointed out that bubbles driven by individual stars in massive and compact star clusters may stall before colliding with neighboring bubbles. In these cases, wind-driven shells stall and fragment. However, the  hot shocked gas produced continuously around individual massive stars continues to expand in the subsonic regime. This enhances the possibility of reaching its closest neighbor unless strong radiative cooling sets in at a distance $R_{\textrm{cool}}$ from the source massive star. If $R_{\textrm{cool}}$ is larger than the half-distance $X$ between neighboring massive stars, hot subsonic blobs finally merge to form a global star cluster wind. However, if in some regions $R_{\textrm{cool}} < X$, the shocked stellar winds cannot merge and expel the residual gas from these zones. Note that the half-distance between neighboring massive stars grows with distance from the star cluster center and in the case of the Gaussian distribution it is:
\begin{equation}\label{half_distance}
X\left(r \right)=R_{c} \left[3 \sqrt{ \pi /2 } \exp \left(0.5 r^{2}/R_{c}^{2} \right)/N_{1G}\right]^{1/3}.
\end{equation}
\begin{figure}
	\includegraphics[width=1.\columnwidth]{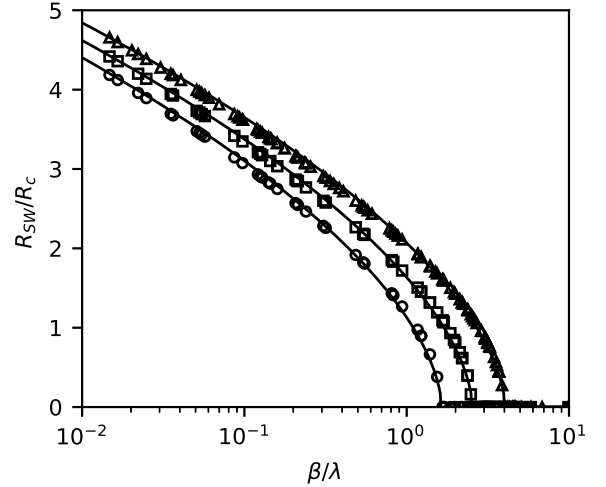}
        \caption{The $R_{\textrm{SW}}/R_c$ ratio as a function of the parameters $\beta/\lambda=\epsilon_{\textrm{1G}}^{1/3}\left(1-\epsilon_{\textrm{1G}} \right)^{1/3} R_{\textrm{c,pc}}^{5/3}M_{\textrm{gas},6}^{-1}$. The triangles, squares and circles symbols present the results of numerical calculations for clouds with metallicities $Z/Z_{\odot}=10^{-2}, 10^{-1}$ and 1, respectively. The analytic fit (equations \ref{WIND_RADIUS}-\ref{WIND_RADIUS2}) to these results is shown by the solid lines.}
    \label{fig3_winds}
\end{figure}
The cooling radius $R_{\textrm{cool}}$ depends on the intra-cloud gas pressure, stellar power, mass loss rate and metallicity (see \citealt{Silich2018, Silich2020}) and should be calculated numerically for the initial conditions discussed in section \ref{clusterModel}. A typical massive star mass loss rate $\log [\dot{M}_{w} (\textrm{M}_{\odot} \textrm{ yr}^{-1})] \approx$ -6.3, -6.2, -5.8  and mechanical luminosity $ \log [L_{w} (\textrm{erg} \textrm{ s}^{-1})] \approx$ 35.5, 35.8, 36.3  were selected from STARBURST99 population synthesis model \citep{1999ApJS..123....3L} upon the assumption that the stellar wind metallicity is $Z/Z_{\odot}=10^{-2}, 10^{-1}$ and 1, respectively. The cooling tables from \cite{Raymond1976} for each one of these metallicities have been used in the calculations, unless otherwise stated.

Fig. \ref{fig2_winds} shows an example of  how the cooling radius $R_{\textrm{cool}}$ (thick solid line) and the mean separation between neighboring massive stars (dotted line) change with distance from the parental cloud center in a $7.14 \times 10^5$ M$_{\odot}$ cloud with $\epsilon_{\textrm{1G}} = 0.1$ and $R_c = 0.87$ pc in the case when the 1G stars and gas metallicity is $Z=10^{-2}$ Z$_{\odot}$. This figure shows that the cooling radius grows faster than the mean separation between neighboring massive stars and therefore it is likely that catastrophic shocked gas cooling and the residual gas large central pressure inhibit the leftover gas expulsion from the central zones, but not from the outskirts of the star-forming cloud. The solid thin vertical line in Fig. \ref{fig2_winds} displays the distance from the cloud center at which $R_{\textrm{cool}} = X$. Hereafter this distance is called the super-wind radius $R_{\textrm{SW}}$.
\begin{figure}
	\includegraphics[width=0.8\columnwidth]{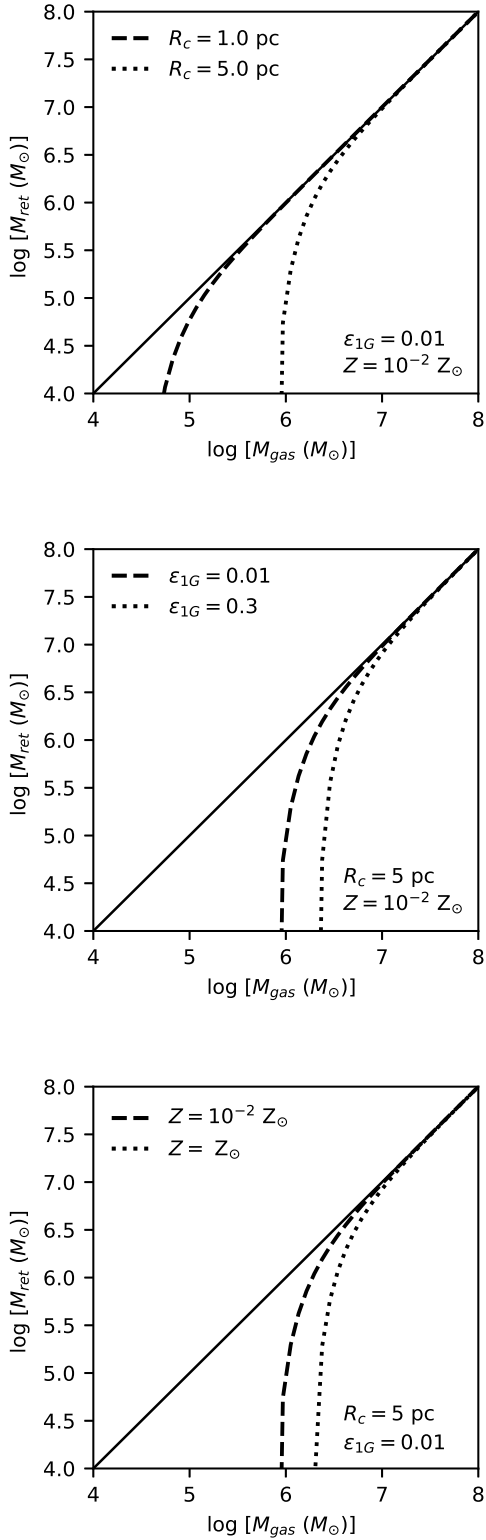}	
    \caption{$M_{\textrm{ret}}$ as a function of the leftover gas mass $M_{\textrm{gas}}$. In the top panel, the star formation efficiency and gas metallicity are fixed (see the inset panel) to present the dependence on the core radius. The middle panel: the same but with fixed $R_{c}$ and $Z$.  Bottom panel: two cases of gas metallicity while the remaining cloud parameters are fixed.}
    \label{mass_retention}
\end{figure}

One can expect $R_{\textrm{SW}}=0$ in low mass and extended clouds, where the central gas pressure is small, but it is nonzero in massive and compact clouds. We estimated numerically the value of $R_{\textrm{SW}}$ for each of the 250 models here considered, with an accuracy of $10^{-3}$ $R_{c}$. The results of these calculations for star-forming clouds with different masses, core radii and star formation efficiencies are shown in Fig. \ref{fig3_winds}. The triangles, squares and circles present cases with metallicities $Z/Z_{\odot}=10^{-2}, 10^{-1}$ and 1, respectively.  This data led us to the approximate relation:
\begin{equation}\label{WIND_RADIUS}
\frac{R_{\textrm{SW}}}{R_{c}}=\left\{
	\begin{array}{ll}
		1.70 \ln \left( 1.02 \beta^{-1} \right)^{0.58},  & \mbox{if }  \beta \leq  1,\\
		0 & \mbox{if } \beta>1.
	\end{array}
\right.
\end{equation}
Here $\beta$ is the $R_{\textrm{cool}}/X$ ratio at the center:
\begin{equation}\label{WIND_RADIUS2}
\beta=\frac{R_{\textrm{cool}} \left(0 \right)}{X\left(0 \right)}=\lambda \epsilon_{\textrm{1G}}^{1/3}\left(1-\epsilon_{\textrm{1G}} \right)^{1/3} \frac{R_{\textrm{c,pc}}^{5/3}}{M_{\textrm{gas},6}},
\end{equation}
where $M_{\textrm{gas},6}=M_{\textrm{gas}}/[10^{6} M_{\odot}]$, $R_{\textrm{c,pc}}=R_{\textrm{c}}/ [1$  $\textrm{pc}]$ and $\lambda$ is a metallicity dependent parameter, which is found to be 0.25, 0.40 and 0.62 for $Z/Z_{\odot}=10^{-2}, 10^{-1}$ and 1, respectively. The Fig. \ref{fig3_winds} presents the superwind radius normalized to the core radius $R_{\textrm{SW}}/R_{\textrm{c}}$ as function of $\beta/\lambda=\epsilon_{\textrm{1G}}^{1/3}\left(1-\epsilon_{\textrm{1G}} \right)^{1/3} R_{\textrm{c,pc}}^{5/3}M_{\textrm{gas},6}^{-1}$. The solid lines are given by equations (\ref{WIND_RADIUS}) and (\ref{WIND_RADIUS2}) while numerical results obtained in calculations with different cloud masses, core radii, star formation efficiencies and metallicities are presented with marks, as indicated in the figure caption. Note that for equal values of $\epsilon_{\textrm{1G}}, R_{\textrm{c}}$ and $M_{\textrm{gas},6}$, the superwind radius is inversely proportional to the gas metallicity. 

Our results differ from those of \citealt{2007ApJ...658.1196T}, \citealt{2007A&A...471..579W}, \citealt{2014ApJ...792..105P}, \citealt{2017ApJ...835...60W}, \citealt{2017MNRAS.470..977L}, as those  papers consider the properties of already formed winds whereas we are after the conditions required to form such winds by measuring the cooling radius of a individual stellar wind and comparing it with the separation between massive stars at different radii.

\subsection{Retained gas mass} \label{gas_retention_section}
The gas mass $M_{\textrm{ret}}$ available for the formation of new stars is located within a sphere with radius $R_{\textrm{SW}}$: $M_{\textrm{ret}}=M_{\textrm{gas}}\left(R_{\textrm{SW}} \right)$. Outside this radius the leftover gas is dispersed by the cluster wind. As an example, Fig. \ref{mass_retention} shows $M_{\textrm{ret}}$ as a function of $M_{\textrm{gas}}$. The solid lines correspond to complete gas retention, i.e., $M_{\textrm{ret}}=M_{\textrm{gas}}$. The top panel presents the case when the core radius is $R_{c}=1$ and 5 pc as dashed and dotted lines, respectively. As shown in the inset panel, the efficiency and gas metallicity are fixed at $\epsilon_{\textrm{1G}}=0.01$ and $Z=10^{-2}$ Z$_{\odot}$. Compact clusters retain a larger fraction of gas because wind-driven bubbles cool more efficiently due to the large gas densities and pressures. Indeed, note that for $R_{c}=5$ pc, the stellar winds remove all the leftover gas of clouds with $\log[ $ M$_{\textrm{gas}} (M_{\odot})]\lesssim 6$. However, for the more compact case of $R_{c}=1$ pc (dashed line), this is only possible for clouds with masses $\log[ $ M$_{\textrm{gas}} (M_{\odot})]\lesssim 4.8$. 

The middle panel of Fig. \ref{mass_retention} shows $M_{\textrm{ret}}$ as function of $M_{\textrm{gas}}$ for $\epsilon_{\textrm{1G}}=0.01$ and 0.3 (dashed and dotted lines) while the core radius and metallicity are fixed (inset panel). Increasing star formation efficiencies lead to a larger number of massive stars (see equation \ref{massive_stars_total}), hence reducing the mean separation $X$ between stars (equation \ref{half_distance}). Therefore, it is easier to have $R_{\textrm{cool}}/X>1$ from inside the cloud and this explain the enhancement of gas expulsion for larger $\epsilon_{\textrm{1G}}$. Note however that the effect of $\epsilon_{\textrm{1G}}$ is not as notorious as the one obtained when the core radius changes (see the top panel). This can be also confirmed by analyzing the dependence on $\epsilon_{\textrm{1G}}$ and $R_{c}$ in equation (\ref{WIND_RADIUS2}). The reason of this limited impact of $\epsilon_{\textrm{1G}}$ on gas expulsion is given by the equation (\ref{turbPfinalINI}). Larger $\epsilon_{\textrm{1G}}$ leads also to larger gas pressures, which reduces the effect of the stellar winds.

Finally, the dashed and dotted lines in the bottom panel of Fig. \ref{mass_retention} are $M_{\textrm{ret}}$ as function of $M_{\textrm{gas}}$ for the metallicities $Z/Z_{\odot}=10^{-2}$ and 1, respectively. The core radius and efficiency are given in the legend. Expectedly, a higher metallicity leads to a stronger feedback as winds from massive stars are more powerful as a function of the gas metallicity. Nevertheless, as before the effect is reduced because gas cooling is also more efficient for increasing metallicities. 

Fig. \ref{mass_retention} shows then that the ability of stellar winds to eject the leftover gas out of the star-forming cloud depends on $M_{\textrm{gas}}$, $R_{c}$, $\epsilon_{\textrm{1G}}$  and $Z$. Our results indicate that the gas mass $M_{\textrm{gas}}$ and core radius $R_{c}$ are the main parameters that determine the retained gas mass $M_{\textrm{ret}}$. Also note that for $M_{\textrm{gas}}\gtrsim 10^{7}$ M$_{\odot}$, the cloud retains most of the gas regardless of the values of the remaining parameters. 

\section{Supernovae feedback}\label{mainEquations}
In massive compact clouds with large intra-cluster gas densities, the cooling rate of individual SNRs is larger than the SN explosion rate, thus hindering their synchronization and overlapping (see appendix \ref{Ap3} for a complete discussion). Hence, here the 3D Thin-Shell approximation \citep[e.g.][]{silich1992,bisno1} is used to follow the evolution of individual SNRs in star-forming clouds. The basic equations of this method are here modified in order to take into account Rayleigh-Taylor (R-T) instabilities.
\subsection{The mass, momentum and energy conservation equations}\label{ejecta_sub}
The SNR is split into a set of Lagrangian elements with indexes $\left(i,j\right)$, where $i \in \{ 1, \ldots, N_{z} \}$ and $j \in \{1, \ldots, N_{\phi}\}$. A grid of $N_{z}=N_{\phi}=40$ is selected for the calculations, which gives a total of $N_{\phi}\left(N_{z}-2 \right)+2=1522$ lagrangian elements for each SNR. The mass and momentum conservation equations for each lagrangian element are then:
\begin{equation}
\frac{d \mu}{dt}=\frac{d \mu_{\textrm{ism}}}{dt}+\frac{d \mu_{\textrm{loss}}}{dt},
\label{fun1a}
\end{equation}
\begin{equation}
\frac{d \mu_{\textrm{ism}}}{dt}=\left\{
	\begin{array}{ll}
		\rho_{g}\left(x,y,x \right) \eta,  & \mbox{if }  P_{\textrm{th}} >  P_{\textrm{g}},\\
		0 & \mbox{if } P_{\textrm{th}} \leq P_{\textrm{g}},
	\end{array}
\right.
\label{fun1}
\end{equation}
\begin{equation}
\frac{d U_{x}}{dt}=\frac{P}{\mu}\frac{\partial \left(y,z \right)}{\partial \left(\lambda_{1}, \lambda_{2} \right)}-\frac{U_{x}}{\mu}\frac{d \mu_{\textrm{ism}}}{dt},
\label{fun2}
\end{equation}
\begin{equation}
\frac{d U_{y}}{dt}=\frac{ P}{\mu}\frac{\partial \left(z,x \right)}{\partial \left(\lambda_{1}, \lambda_{2} \right)}-\frac{U_{y}}{\mu}\frac{d \mu_{\textrm{ism}}}{dt},
\label{fun3}
\end{equation}

\begin{equation}
\frac{d U_{z}}{dt}=\frac{P}{\mu}\frac{\partial \left(x,y \right)}{\partial \left(\lambda_{1}, \lambda_{2} \right)}-\frac{U_{z}}{\mu}\frac{d \mu_{\textrm{ism}}}{dt},
\label{fun4}
\end{equation}
\begin{equation}
P=\left\{
	\begin{array}{ll}
		P_{\textrm{th}},  & \mbox{if }  P_{\textrm{th}} >  P_{\textrm{g}},\\
		P_{\textrm{th}}-P_{\textrm{g}} & \mbox{if } P_{\textrm{th}} \leq  P_{\textrm{g}},
	\end{array}
\right.
\label{fun5}
\end{equation}

\begin{equation}
\frac{d \mathbf{r}}{dt}=\mathbf{V_{\textrm{shock}}},
\label{fun6}
\end{equation}
\begin{equation}\label{fun7}
\mathbf{V}=\left\{
	\begin{array}{ll}
		\frac{\gamma+1}{2}\mathbf{U},  & \mbox{if }  t \leq  t_{\textrm{cool}},\\
		\mathbf{U} & \mbox{if } t > t_{\textrm{cool}},
	\end{array}
\right.
\end{equation}
\noindent  where $\mu$, $x,y,z, U_{x}, U_{y}, U_{z}$ are the mass, positions and velocities of the lagrangian element, $\gamma=5/3$ is the specific heats ratio,  $P_{\textrm{g}}$ and $\rho_{g}\left(x,y,z \right)$ are the ambient gas pressure and density, $\partial \left(x_{i},x_{j} \right)/\partial \left(\lambda_{1}, \lambda_{2} \right)$ are the jacobians, $d \mu_{\textrm{ism}}/dt$ is the rate of mass swept from the ambient gas and $d \mu_{\textrm{loss}}/dt$ is the mass loss rate due to RT instabilities (see section \ref{RT_loss}), respectively. The term $\mathbf{V}$ is the leading shock velocity, $t_{\textrm{cool}}$ is the transition time from the adiabatic to the radiative stage, $P_{th}$ is the total thermal pressure and:
\begin{dmath}
\eta=V_{x}\frac{\partial \left(y,z \right)}{\partial \left(\lambda_{1}, \lambda_{2} \right)}+V_{y}\frac{\partial \left(z,x \right)}{\partial \left(\lambda_{1}, \lambda_{2} \right)}+V_{z}\frac{\partial \left(x,y \right)}{\partial \left(\lambda_{1}, \lambda_{2} \right)}.
\end{dmath}
The jacobians are calculated following \cite{bisno1} and the transition time $t_{\textrm{cool}}$ is calculated as in  \cite{2019MNRAS.488..978J}.\\
\indent Equation (\ref{fun1}) states that a Lagrangian element accumulates mass only when the inner thermal pressure is larger than the external gas pressure. Equations (\ref{fun2}-\ref{fun4}) take into account the change in the SNR momentum due to the mass loss via RT instabilities. \\
\indent The thermal pressure inside the remnant is determined by the equation:
\begin{equation}\label{thermalP}
P_{\textrm{th}}=\left(\gamma-1 \right) \frac{E_{\textrm{th}}}{\Omega-\left(4\pi/3 \right)R_{RS}^{3}},
\end{equation}
where $\Omega$ is the SNR volume:
\begin{equation}
\Omega=\frac{1}{3} \int \int \left[x \frac{\partial \left(y,z \right)}{\partial \left(\lambda_{1}, \lambda_{2} \right)} +y\frac{\partial \left(z,x \right)}{\partial \left(\lambda_{1}, \lambda_{2} \right)} +z\frac{\partial \left(x,y \right)}{\partial \left(\lambda_{1}, \lambda_{2} \right)}\right] d \lambda_{1} d \lambda_{2},
\label{surf4}
\end{equation} 
\begin{figure*}
	\includegraphics[width=1.5\columnwidth]{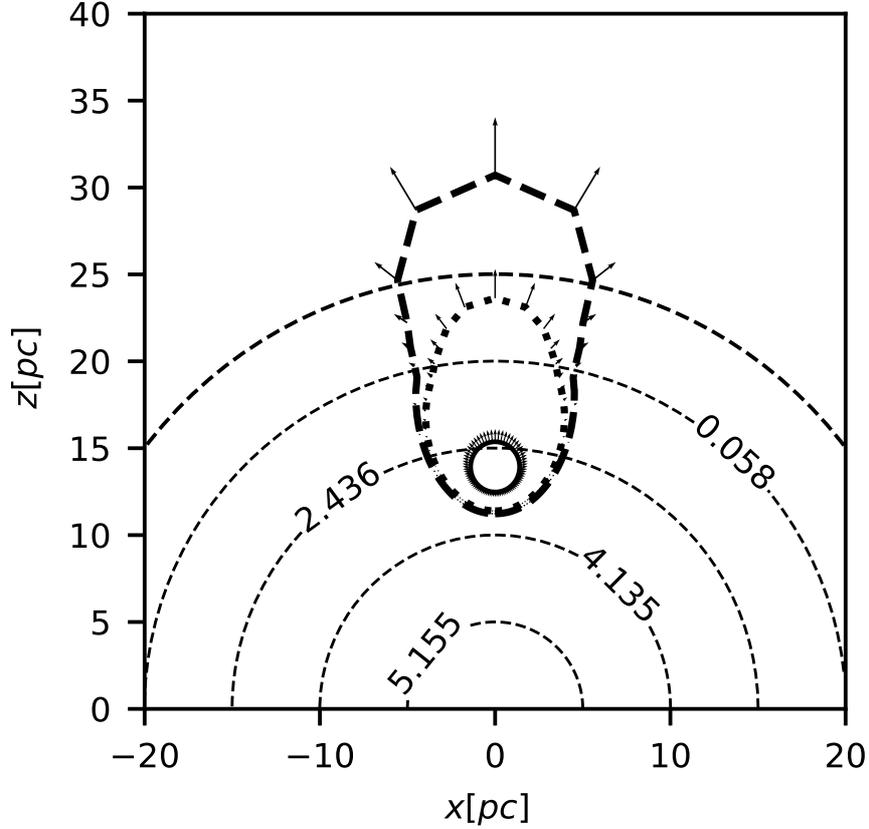}
        \caption{Blowout, a typical evolution. Slices of a SNR on the $x$-$z$ plane. The ambient gas density is depicted by the contour lines, with the log of the number density (cm$^{-3}$) indicated in each line. The last circular line is the cloud boundary with radius $R_{SC}$. The solid, dotted and dashed contours
correspond to $1.09 \times 10^{3}$ yr, $12.2\times 10^{3}$ yr and $15.3\times 10^{3}$ yr after
the explosion. The velocity vectors are shown for each snapshot to indicate the flow motion. These vectors are normalized with respect to the largest velocity of the lagrangian elements ($\approx 2.44 \times 10^{3}$ km s$^{-1}$) at the end of the simulation, which in this case is the north pole velocity (the uppermost vector).}
    \label{fig1}
\end{figure*}
$E_{\textrm{th}}$ is the total thermal energy and $R_{RS}$ is the position of the reverse shock. \\
\indent The evolution of the SNR energy is given by the energy conservation equation:
\begin{dmath}\label{thermalEa}
\frac{d}{dt}\left(E_{\textrm{th}}+E_{\textrm{k,free}}+E_{\textrm{k,ej}}+E_{\textrm{k,s}}\right)=-Q,
\end{dmath}
where $E_{\textrm{k,free}}$ and $E_{\textrm{k,ej}}$ are the kinetic energies of the free and shocked 
ejecta gas, respectively. $E_{\textrm{k,s}}=0.5 \sum_{i,j} \mu U^{2}$ is the kinetic energy of the 
shocked ambient gas shell, $Q$ is the energy loss due to radiative cooling:
\begin{equation}
Q=n_{e}n_{i}\Lambda \left(T,Z\right),
\label{cool_function}
\end{equation} 
where $\Lambda \left(T,Z\right)$ is the cooling function for a metallicity $Z$, and $n_{i}\approx n_{e}$ are the ion and electron number densities of the cooling gas. \\
\indent Equation (\ref{thermalEa}) leads to a differential equation for the total thermal energy:
\begin{dmath}\label{thermalE}
\frac{d E_{\textrm{th}}}{dt}=-\frac{d E_{\textrm{k,free}}}{dt}-\frac{d E_{\textrm{k,ej}}}{dt}- \frac{d E_{\textrm{k,s}}}{dt}-Q.
\end{dmath}
The SNR is assumed to begin in the free-expansion stage (see Appendix \ref{Ap1}), with a $r^{-2}$ density profile for the ejecta with total mass $M_{ej}=3 M_{\odot}$ and total energy $E_{0}=10^{51}$ erg. The kinetic energies as well as the position of the reverse shock and the gas cooling $Q$ are calculated following \cite{2019MNRAS.488..978J}.
%---------------------------------------------------------------------------

%---------------------------------------------------------------------------
\subsection{The Rayleigh-Taylor instability}
The RT time-scale is given by \citep[e.g.][]{2013A&A...557A.140B}:
\begin{equation}\label{eqtau}
\tau^{2}=\Delta R\left[\left( \frac{dU_{x}}{dt}\right)^{2}+\left( \frac{dU_{y}}{dt}\right)^{2}+\left( \frac{dU_{z}}{dt}\right)^{2}\right]^{-1/2},
\end{equation}
where $\Delta R$ is the shell thickness:
\begin{equation}\label{eqDR}
\Delta R=\frac{M}{d \Sigma \rho_{\textrm{shell}}}=\frac{M k T_{\textrm{shell}}}{d \Sigma \mu P_{th}}.
\end{equation}
In equation (\ref{eqDR}), $\rho_{\textrm{shell}}$ is the shell density and $d \Sigma$ the surface area of the Lagrangian element:
\begin{dmath}
d \Sigma^{2}=\left(\frac{\partial \left(y,z \right)}{\partial \left(\lambda_{1}, \lambda_{2} \right)} \right)^{2}+\left(\frac{\partial \left(z,x \right)}{\partial \left(\lambda_{1}, \lambda_{2} \right)} \right)^{2}+\left(\frac{\partial \left(x,y \right)}{\partial \left(\lambda_{1}, \lambda_{2} \right)} \right)^{2}.
\label{surf2}
\end{dmath}
The temperature of the Lagrangian element $T_{\textrm{shell}}$ at the adiabatic stage is calculated from the Rankine-Hugoniot relations \citep[e.g.][]{draine1993theory}  and is assumed to have a constant value at the radiative stage:
\begin{equation}\label{eq:A1}
T_{\textrm{shell}}=\left\{
	\begin{array}{ll}
		1.38 \times 10^{7}\left(\frac{V}{1000 \textrm{ km s}^{-1}} \right)^{2} \textrm{ K,} & \mbox{if }  t<t_{\textrm{cool}},\\
		10^{4}\textrm{ K,} & \mbox{if } t > t_{\textrm{cool}},
	\end{array}
\right.
\end{equation}

\subsection{Mass loss during the RT instability}\label{RT_loss}
The onset of the RT instability occurs at $t_{ac}$, which for a given Lagrangian element is determined by the condition:
\begin{equation}\label{eqacelera}
\mathbf{n} \cdot \frac{d \mathbf{U}}{dt} = \frac{\partial \left(y,z \right)}{\partial \left(\lambda_{1}, \lambda_{2} \right)}\frac{dU_{x}}{dt}+\frac{\partial \left(z,x \right)}{\partial \left(\lambda_{1}, \lambda_{2} \right)}\frac{dU_{y}}{dt}+\frac{\partial \left(x,y \right)}{\partial \left(\lambda_{1}, \lambda_{2} \right)}\frac{dU_{z}}{dt}>0,
\end{equation}
where $\mathbf{n}$ is the normal vector of the Lagrangian element:
\begin{equation}
\mathbf{n} =\frac{1}{d \Sigma}\left[ \frac{\partial \left(y,z \right)}{\partial \left(\lambda_{1}, \lambda_{2} \right)} \mathbf{i}+\frac{\partial \left(z,x \right)}{\partial \left(\lambda_{1}, \lambda_{2} \right)} \mathbf{j}+\frac{\partial \left(x,y \right)}{\partial \left(\lambda_{1}, \lambda_{2} \right)} \mathbf{k} \right],
\end{equation} 
It is assumed that a RT unstable Lagrangian element fragments at the time $t_{RT}$ such that \citep[e.g.][]{2013A&A...557A.140B}:
\begin{equation}
\left(t_{RT}-t_{ac}\right)>3 \tau\left(t_{RT} \right).
\end{equation}
After this time, the hot gas escapes from the SNR interior through the fragmented shell and follows the accelerated leading shock. This process can be modeled by removing the accumulated mass from the RT unstable Lagrangian element. Here, it is assumed that a RT unstable Lagrangian element loses its mass as (e.g. \citealt{1998MNRAS.299..249S}):
\begin{equation}\label{massloss}
\frac{d M_{\textrm{loss}}}{dt}=\left\{
	\begin{array}{ll}
		-\frac{M_{RT}}{\tau_{RT}}\exp \left[-\left(\frac{t-t_{RT}}{\tau_{RT}}\right) \right],  & \mbox{if }  t>t_{\textrm{RT}},\\
		0 & \mbox{if } t < t_{\textrm{RT}},
	\end{array}
\right.
\end{equation}
where $M_{RT}=M\left(t_{RT} \right)$ and $\tau_{RT}=\tau \left(t_{RT}\right)$ are the mass and 
the RT timescale at the beginning of the RT instability.
\begin{figure}
	\includegraphics[width=\columnwidth]{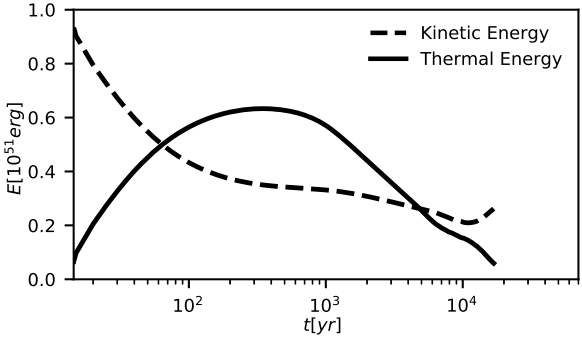}
    \caption{The SNR kinetic (dashed line) and thermal (solid) energies.}
    \label{fig2}
\end{figure}

\begin{figure}
	\includegraphics[width=\columnwidth]{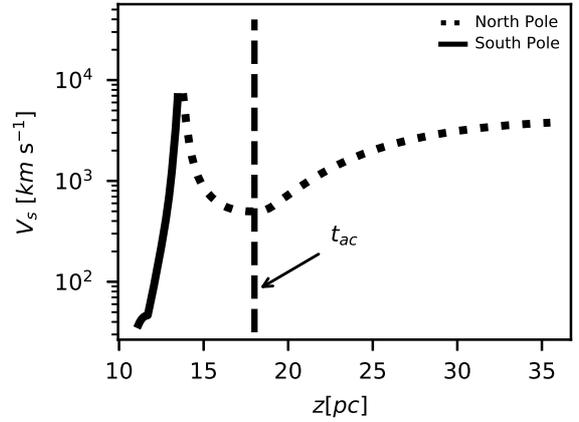}
    \caption{The shock velocity as a function of the $z$ coordinate. The solild line is the south pole velocity and the dotted line is the north pole velocity. The dashed vertical line indicates the acceleration time $t_{ac}$.}
    \label{fig3}
\end{figure}
\begin{table*}
\caption{From the left to the right columns: model ID, the 1G mass $M_{\textrm{1G}}$, the star-formation efficiency $\epsilon_{\textrm{1G}}$, the total mass $M_{\textrm{tot}}$, the leftover gas mass $M_{\textrm{gas}}$ , the core radius $R_{c}$, the gas central density $n_{0}$, the cloud central gas pressure $P_{0}$ and the cloud metallicity.}
\begin{center}
\begin{tabular}{c c c c c c c c c}
\toprule 
Model& $M_{\textrm{1G}}$  & $\epsilon_{\textrm{1G}}$ & $M_{\textrm{tot}}$ & $M_{\textrm{gas}}$ &  $R_{c}$& $n_{0}$ &$P_{\textrm{0}}$& $Z/Z_{\odot}$ \\
\toprule 
& $M_{\odot}$ & &  $M_{\odot}$ &$M_{\odot}$ & pc & cm$^{-3}$ & dyn cm$^{-2}$&  \\
\toprule 
\toprule 
1&2.0e+05&0.01&2.00e+07&1.98e+07&1.5&1.18e+07&2.5e-03&$10^{-2}$\\
2&2.0e+05&0.01&2.00e+07&1.98e+07&4.0&6.25e+05&4.9e-05&$10^{-2}$\\
3&2.0e+05&0.025&8.00e+06&7.80e+06&1.5&4.67e+06&3.9e-04&$10^{-2}$\\
4&2.0e+05&0.025&8.00e+06&7.80e+06&4.0&2.46e+05&7.7e-06&$10^{-2}$\\
5&2.0e+05&0.07&2.86e+06&2.66e+06&1.5&1.59e+06&4.7e-05&$10^{-1}$\\
6&2.0e+05&0.07&2.86e+06&2.66e+06&4.0&8.38e+04&9.3e-07&$10^{-1}$\\
7&2.0e+05&0.1&2.00e+06&1.80e+06&1.5&1.08e+06&2.2e-05&$10^{-1}$\\
8&2.0e+05&0.1&2.00e+06&1.80e+06&4.0&5.68e+04&4.4e-07&$10^{-1}$\\
9&2.0e+05&0.1&2.00e+06&1.80e+06&1.5&1.08e+06&2.2e-05&1.0\\
10&2.0e+05&0.1&2.00e+06&1.80e+06&4.0&5.68e+04&4.4e-07&1.0\\
11&2.0e+05&0.25&8.00e+05&6.00e+05&1.5&3.59e+05&3.0e-06&1.0\\
12&2.0e+05&0.25&8.00e+05&6.00e+05&4.0&1.89e+04&5.9e-08&1.0\\
\toprule 
\toprule
\end{tabular}
\end{center}
\label{tabla1}
\end{table*}
\section{SN blowout: a typical evolution}\label{typicalBlowout}
A three zone structure forms in the ISM when the high velocity supernova ejecta collides with the circumstellar gas. There is first an outer shell where all the gas  swept up by the leading shock is continuously accumulated. Interior to this is the SN ejecta, with an outer region filled with all the ejected gas that has already being thermalized at a reverse shock, and a central zone with the still un-shocked ejecta. The shocked ejecta is separated from the ambient swept up gas by a contact discontinuity while a reverse shock, driven towards the explosion center, rapidly manages to  overtake and thermalized all the ejecta. The large pressure attained by the shocked ejecta is then able to push the swept up gas and the leading shock away from the explosion center while the interior hot gas becomes rarefied while filling a larger volume.

This structure evolves with time and is known as a supernova remnant (SNR). Its evolutionary sequence includes the thermalisation of the ejecta, the quasi adiabatic Sedov-Taylor (ST) stage, the so called snowplough stage and the final phase in which the expansion proceeds by conservation of the momentum gathered through the evolution (\cite{1977Chevalier}, and see extensive reviews by \citealt{astrowaves, bisno1}). At the initial stage the leading and the reverse shocks form and the kinetic energy of the ejecta is transformed into thermal energy of the shocked ejecta and into the thermal and kinetic energies of the swept up shell \citep[e.g.][]{chevalier1982self,hamilton1984new, chevalier1984interaction,1999Mckee}. The ST stage begins when the reverse shock reaches  the explosion site (\citealt{1946RSPSA.186..273T, 1959sdmm.book.....S}). This phase is characterized by negligible radiative losses of energy and if the evolution proceeds in a homogeneous ambient medium, the hydrodynamic variables and expansion radius follow a self-similar solution. The snowplough stage begins when strong radiative cooling sets in within the swept up ambient gas. At this stage the outer shell collapses into a dense, cold and thin structure, while the shocked ejecta remains hot holding its high thermal pressure \citep[e.g.][]{cioffi1988dynamics,Blondin1998}. At the final, momentum conserving phase, the shocked ejecta also cools down by radiation, while the outer shell of swept up matter slows down in a momentum conservation regime \citep[e.g.][]{1999Mckee,2017Tang}. Note that in remnants evolving in a high density medium ($n > 10^4$ cm$^{-3}$) the shocked ejecta and the outer shell cool so rapidly that remnants avoid the ST stage \citep[e.g.][]{ terlevich1992starburst, 2019MNRAS.488..978J}.

Remnants evolving in a cloud or a medium with a strong density gradient become strongly distorted. The section of the leading shock and that of the shell of swept up matter, facing away from the cloud center begin to accelerate after reaching a few times the core radius of the gas density distribution. This promotes the development of Rayleigh-Taylor instabilities and the fragmentation of the accelerated shell sections which allow the hot thermalized SN ejecta to stream  between  shell fragments and out of the SNR interior. The hot ejecta follows the accelerated shock as this moves into even lower densities while the remains of the shell, holding most of the swept up matter, but having lost its driving pressure, rapidly slow down to finally remain within the cloud. The whole process is known in the literature as a blowout event.

Figs. \ref{fig1}-\ref{fig3} illustrate a blowout event. It was assumed that the explosion occurs within a cloud with $M_{\textrm{gas}} = (1 - \epsilon_{\textrm{1G}}) M_{\textrm{tot}} = 9.9 \times 10^6$ M$_{\odot}$, $R_c = 4.0$ pc and $R_{\textrm{SC}} = 25$ pc. The exploding star was located at the distance $z_0 = 13.6$ pc from the cloud center where the gas density is $0.96 \times 10^3$ cm$^{-3}$. Fig. \ref{fig1} presents slices of the remnant in the x-z plane for the post-explosion times indicated in the figure caption. The arrows show the velocities of the lagrangian elements for each snapshot. For remnants evolving in a constant density medium, during the ST stage, their total kinetic and thermal energies hold approximately constant values (0.3 $E_0$ and 0.7 $E_0$, respectively; where $E_0$ is the explosion energy). This clearly does not occur here, as shown in Fig. \ref{fig2}, which presents the SNR kinetic (dashed line) and thermal (solid line) energies. This figure shows that the SNR evolution in a dense cloud differs drastically from the classic ST stage as radiative cooling affects the SNR energy balance significantly. Finally, Fig. \ref{fig3} shows the SNR north (dashed line) and south (solid line) pole velocities as a function of the z-coordinate. Here the dashed vertical line marks the beginning of the acceleration phase ($t_{\textrm{ac}}$).

Initially, the SNR is approximately spherical (see Fig. \ref{fig1}) as at early
times the density gradient is not important. At this stage the expansion
velocity is large ($\approx 10^{4}$ km s$^{-1}$). However, the SNR evolution
speeds-up in a high ambient density medium. Indeed, the ejecta kinetic energy, that at early times accounts for most of the remnant 
energy, is rapidly transformed into  thermal energy reaching a maximum
value at about 400 yr after the SN explosion. It then decreases rapidly due to
strong radiative cooling (see Fig. \ref{fig2}). Therefore in a high density
media SNRs completely avoid the quasi-adiabatic Sedov-Taylor phase
\citep{terlevich1992starburst,2019MNRAS.488..978J}.

The SNR north pole begins to accelerate at $t\approx 5.98 \times 10^{3}$ yr when $z_{\textrm{north}}\approx 18.0$ pc, while the south pole continues to decelerate (see Fig. \ref{fig3}).
Nevertheless, the north pole cooling time is small because of the large gas density. This results in a short RT time
scale as the radiative shell thickness is small (see equations \ref{eqtau}
and \ref{eqDR}). Therefore, soon the north pole becomes RT unstable  and the
SNR shell breaks-out. The inner hot gas, which by this time has an average temperature of $3.7 \times 10^{8}$ K and number density of $ \approx 1$ cm$^{-3}$, then flows away of the remnant and follows the shock as this
accelerates into a decreasing gas density. (see Figs. \ref{fig1} and \ref{fig3}). 

The SN blowout is evident at the age of $15.3\times 10^{3}$ yr. The
northern sections of a new shell formed around the fragmented sections of the SNR,
move with large velocity out of the parental cloud whereas the southern ones decelerate to velocities which are smaller than the intra-cloud gas turbulent speed. The shock at the southern part of the remnant then vanishes whereas the thermalized ejecta enriched with the iron-group elements is vented  into the ambient ISM.

\begin{figure*}
	\includegraphics[width=1.6\columnwidth]{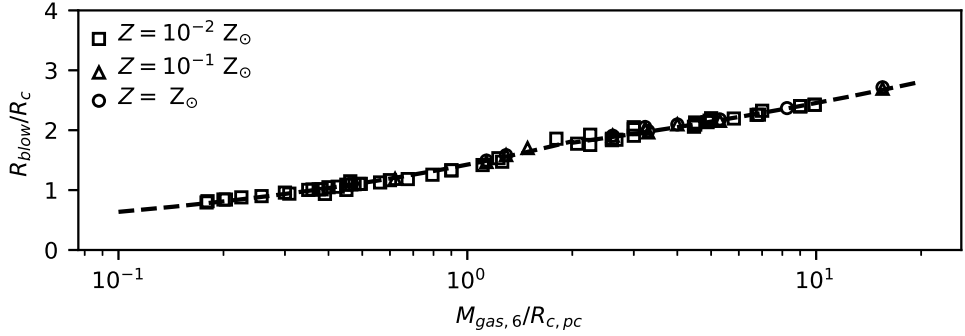}	
    \caption{The blowout radii as a function of the gas concentration $M_{\textrm{gas},6}/R_{\textrm{c,pc}}$ for a large number of models with different masses, core radii and star formation efficiencies, including those presented in Table \ref{tabla1}. Squares, triangles and circles are models with metallicity $Z/Z_{\odot}=10^{-2},10^{-1}$ and 1, respectively. The dashed line is the best fit to the data (see equation \ref{blow_function}). }
    \label{fig4}
\end{figure*}

The SNRs blowout occurs  because the gas density drops rapidly outside of the central zone of the cloud.  However, if the SN explosion occurs close to the cloud center, the SNRs rapidly lose most of their initial energy via catastrophic cooling and this may inhibit their blowout (see Appendix \ref{Ap3}). Such remnants remain pressure confined in the central zone of the cluster and their SN products may eventually mix with the leftover gas. For a given cloud, a critical radius $R_{\textrm{blow}}$, hereafter called the blowout radius, splits cases with and without SNR confinement. This blowout radius allows one to estimate the number of supernovae trapped within the cluster in each particular case. 

\section{The blowout radius}\label{blowradii_sec}
We have calculated $R_{\textrm{blow}}$ for a set of 75 cloud models with different gas masses, star formation efficiencies, core radii and metallicities (the set of input parameters for 12 of these models are presented in Table \ref{tabla1}).  For each model, this was done by varying the explosion location $z_{0}$, starting from values close to the cloud center, and moving outwards in steps of $0.025 R_{c}$. $R_{\textrm{blow}}$ is the minimum value of $z_{0}$ for which the north pole of the SNR is able to accelerate and thus lead to blowout. Remnants that explode within these radii stall in the central zones of the cluster and are assumed to disperse and contaminate the gas left over from star formation. \\
\indent The blowout radii as a function of the gas cloud concentration $M_{\textrm{gas,6}}/R_{\textrm{c,pc}}$ \citep[e.g.][]{ 2018ARA&A..56...83B, 2019A&A...624A..24C} are shown in Fig \ref{fig4} by square, triangle and circle symbols. The numerical results are best fitted by the power-law (dashed
line):
%------------------------------------------------------------------------
\begin{equation}
\frac{R_{\textrm{blow}}}{R_{c}}=\left\{
	\begin{array}{ll}
		1.4246 \left(\frac{M_{\textrm{gas,6}}}{R_{\textrm{c,pc}}} \right)^{0.35},  & \mbox{if    }  \frac{M_{\textrm{gas,6}}}{R_{\textrm{c,pc}}} \leq 2,\\
		1.5639\left(\frac{M_{\textrm{gas,6}}}{R_{\textrm{c,pc}}} \right)^{0.2},  & \mbox{if    }  \frac{M_{\textrm{gas,6}}}{R_{\textrm{c,pc}}} > 2,
	\end{array}
\right.
\label{blow_function}
\end{equation}
%--------------------------------------------------------------------------
Note that the blowout radius is a function of the gas mass and core radius but not of the gas metallicity, as in high density environments (see column 7 of Table \ref{tabla1}), the early SNR evolution is determined basically by free-free cooling, which does  not depend, as line radiative cooling, on  the gas metallicity (see \citealt{2019MNRAS.488..978J}). 

The blowout radii and the iron yield per supernova allow one to estimate the metallicity enhancement provided by the 1G stars for each model, as discussed in the following sections. However, equation (\ref{blow_function}) was derived under the assumption that SNRs evolve within the gas left over from star formation ($R_{\textrm{SW}}>0$). On the other hand, all the SN ejecta is assumed to be expelled from the cluster as part of a global cluster wind in clouds with $R_{\textrm{SW}}$ = 0.
%------------------------------------------------------------------------------

{%\renewcommand{\arraystretch}{1.8}

\section{Chemical enrichment of the leftover gas}\label{enrichment_section}
\begin{figure*}
	\includegraphics[width=2\columnwidth]{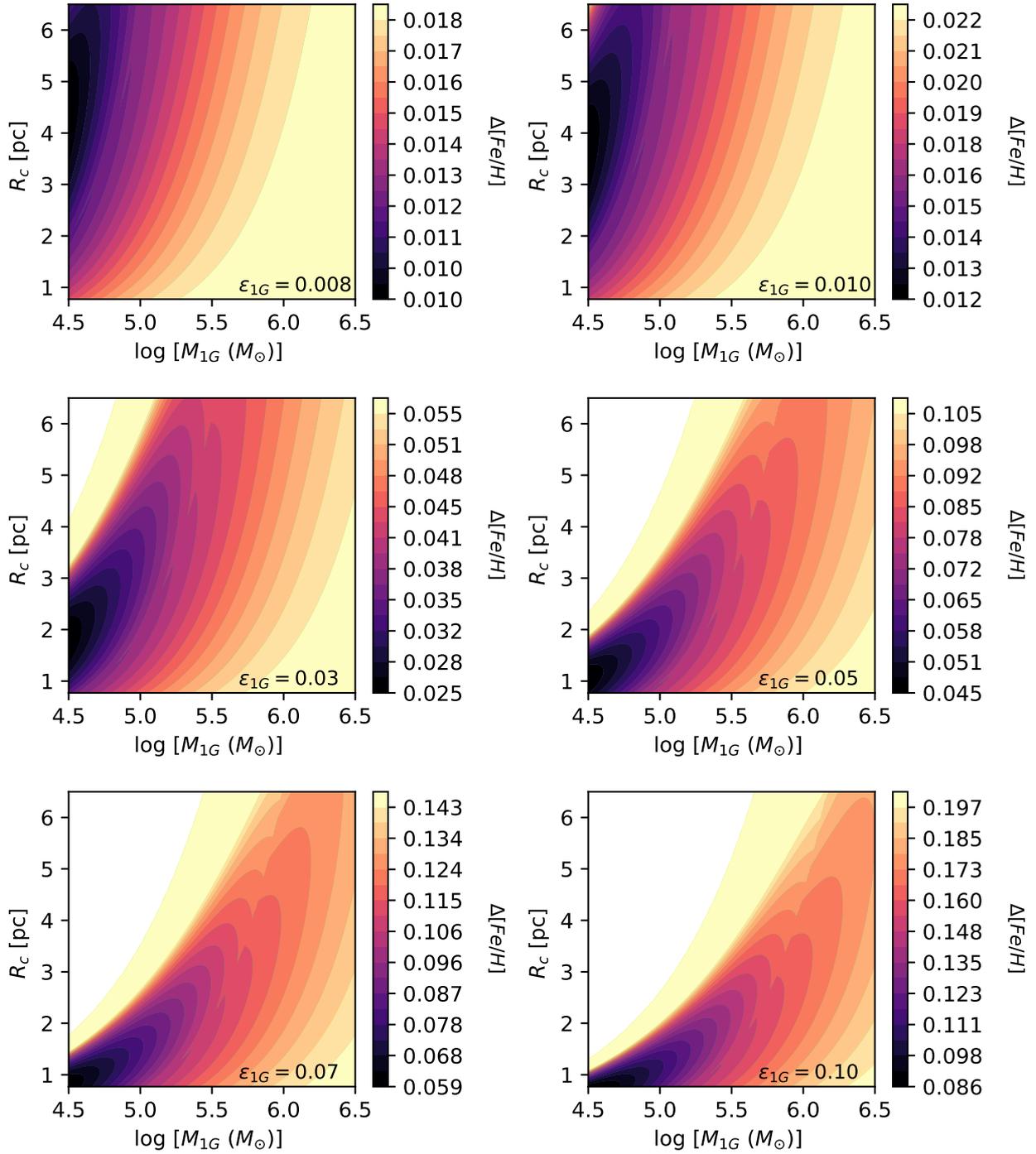}
    \caption{The panels indicate the large variety of initial conditions here used for clouds that form a first stellar generation with mass $M_{\textrm{1G}}$ and core radius $R_c$ for different values of the star formation efficiency ($\epsilon_{\textrm{1G}}$, shown in every panel) and for the metallicity $Z=10^{-1}$ Z$_{\odot}$. For each of these cases the base of the cluster wind ($R_{\textrm{SW}}$; equation \ref{WIND_RADIUS}) and the blowout radius ($R_{\textrm{blow}}$; equation \ref{blow_function}) were determined. The cloud mass below $R_{\textrm{SW}}$ was assumed to throughly mix with the metals produced by the total number of trapped supernovae. The colours indicate the resultant metallicity enhancement  ($\Delta [\textrm{Fe/H}]$) for every case.
 The white areas in every plot, indicate cases for which the base of the cluster wind is at the center of the considered cloud and thus causes a global cluster wind and with it, a total dispersal of the gas left over after star formation. Such cases, in our scenario, do not lead to a second stellar generation.  
Note that  $\Delta$[Fe/H] spans through larger values for increasingly larger values of $\epsilon_{\textrm{1G}}$. }
    \label{Fe_increase}
\end{figure*}
\begin{figure*}
	\includegraphics[width=2\columnwidth]{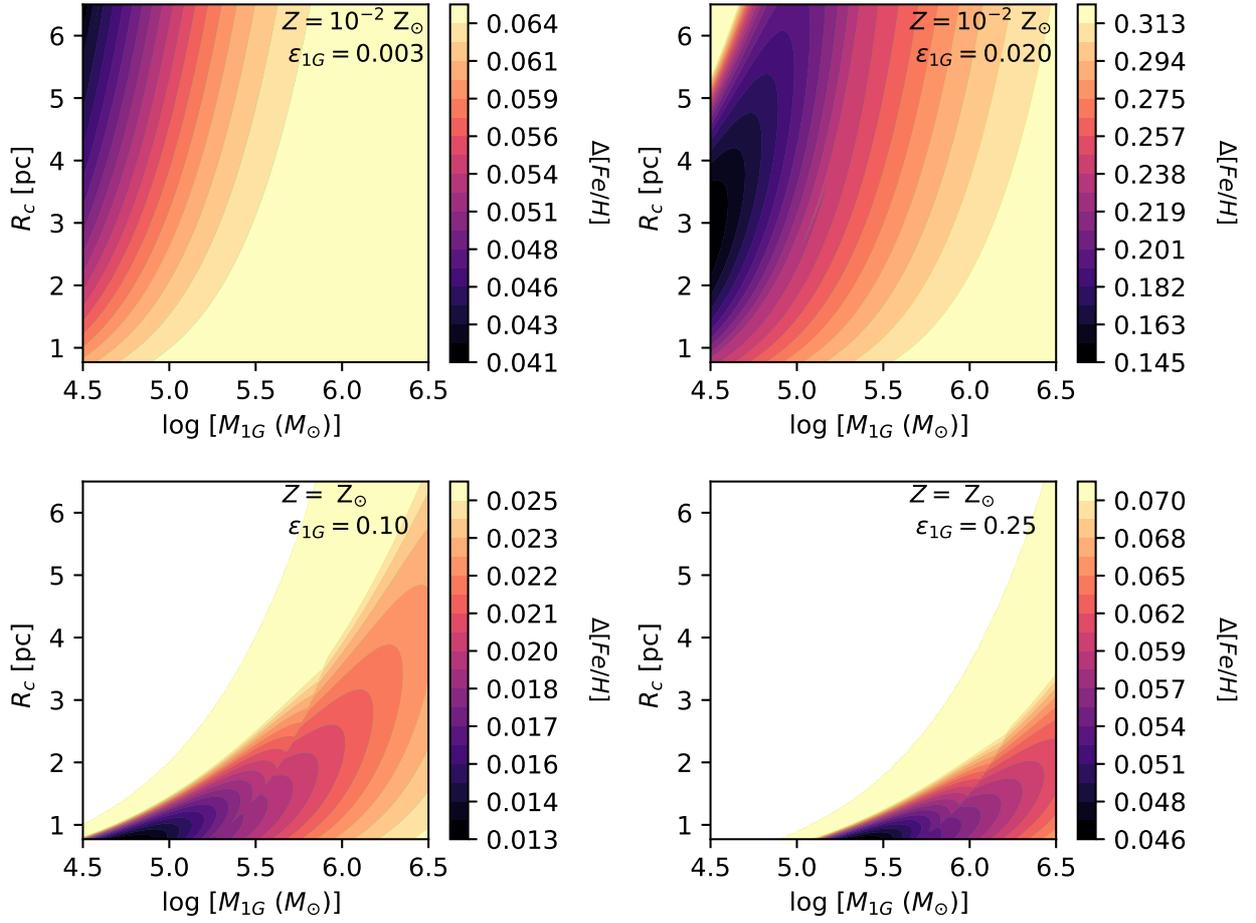}
    \caption{Same as Fig \ref{Fe_increase} but each row present two cases with different star formation efficiency for the metallicities $Z=10^{-2}$ Z$_{\odot}$ and $Z=$ Z$_{\odot}$, respectively. }
    \label{Fe_increase_b}
\end{figure*}
\begin{table*}
\caption{From left to right the columns display: model ID, the total number of stars with masses $>8 M_{\odot}$, the super-wind radius  normalized to the core radius $R_{\textrm{SW}}/R_{c}$, the blowout radius $R_{\textrm{blow}}/R_{c}$, the number of trapped supernovae $N_{\textrm{trapped}}$, the mass of iron deposited by these supernovae $M_{\textrm{TSN}}$, the increase of the iron metallicity from the primordial value to that retained by the left over cloud after contamination by the trapped SN:  $\Delta[\textrm{Fe/H}]=[\textrm{Fe/H}]_{2}-[\textrm{Fe/H}]_{1}$ and the total mass of retained leftover gas $M_{\textrm{ret}}$. See the text for a discussion of these values.}
\begin{center}
\begin{tabular}{c c c c c c c c }
\toprule 
Model&$N_{>8 M_{\odot}}$&$R_{\textrm{SW}}/R_{c}$&$R_{\textrm{blow}}/R_{c}$&$N_{\textrm{trapped}}$& $M_{\textrm{TSN}}$ & $ \Delta [\textrm{Fe/H}]$ &  $M_{\textrm{ret}}$ \\
\toprule 
&  & & && $M_{\odot}$ &dex&$M_{\odot}$\\
\toprule 
\toprule 
1&2.00e+03&4.477&2.59&1.84e+03&1.286e+02&0.176&1.980e+07\\
2&2.00e+03&3.6&2.138&1.59e+03&1.112e+02&0.157&1.971e+07\\
3&2.00e+03&3.829&2.159&1.60e+03&1.122e+02&0.324&7.783e+06\\
4&2.00e+03&2.822&1.799&1.29e+03&9.011e+01&0.286&7.435e+06\\
5&2.00e+03&2.656&1.74&1.22e+03&8.577e+01&0.103&2.471e+06\\
6&2.00e+03&1.148&1.235&6.46e+02&4.526e+01&0.169&7.314e+05\\
7&2.00e+03&2.277&1.518&9.77e+02&6.839e+01&0.13&1.514e+06\\
8&2.00e+03&0.0&-&-&-&-&0.000e+00\\
9&2.00e+03&1.903&1.518&9.77e+02&6.839e+01&0.018&1.250e+06\\
10&2.00e+03&0.0&-&-&-&-&0.000e+00\\
11&2.00e+03&0.0&-&-&-&-&0.000e+00\\
12&2.00e+03&0.0&-&-&-&-&0.000e+00\\
\toprule 
\toprule
\end{tabular}
\end{center}
\label{tabla2}
\end{table*}
Heavy elements, such as the iron injected by supernovae, when trapped within the blowout radius, are a source able to pollute the residual gas from which a 2G of stars are likely to form. Let us name the iron mass returned by these trapped supernovae as $M_{\textrm{TSN}}$. The enriched gas metallicity $Z_1$ is then determined by the equation
%---------------------------------------------------------------------------
\begin{equation}\label{first_metal_eq}
M_{\textrm{ret}} Z_{1}  =  M_{\textrm{TSN}}  +   M_{\textrm{ret}}  Z_0 ,
\end{equation}
%--------------------------------------------------------------------------
where $Z_{0}$ is the primordial gas metallicity and $M_{\textrm{ret}}$ is the retained gas mass (section \ref{gas_retention_section}). Note that $Z_{0}  M_{\textrm{ret}}+M_{\textrm{TSN}}$ is the total amount
of iron in the enriched gas. Then: 
\begin{equation}
 Z_{1}  =   \frac {M_{\textrm{TSN}}} {M_{\textrm{ret}}}  +   Z_0 .
\end{equation}
The gas metallicity may be also presented in solar units:
\begin{equation}\label{metallicity_difference}
[\textrm{Fe/H}]_{1}  = \log \left(10^{[\textrm{Fe/H}]_{0}} + \frac {M_{\textrm{TSN}}} {Z^{fe}_\odot  M_{\textrm{ret}}} \right),
\end{equation}
%--------------------------------------------------------------------------
where $Z^{fe}_\odot=0.0013$ \citep{2008MNRAS.391..354R}.  

One can obtain now the enhancement of the residual gas metallicity provided by trapped SNe by making use of equations (\ref{blow_function}-\ref{first_metal_eq}) and (\ref{metallicity_difference}). The trapped iron mass is calculated upon the assumption that each SN injects $M_{\textrm{Fe}} = 0.07$ M$_{\odot}$ of iron \citep[e.g.][]{2003ApJ...582..905H,2008MNRAS.391..354R}. 

The iron spread $\Delta [\textrm{Fe/H}]$ as a function of the 1G stellar mass $M_{\textrm{1G}}$, $R_c$ and the 1G star formation efficiency $\epsilon_{\textrm{1G}}$ in the case when the initial gas metallicity is $Z=10^{-1}$ Z$_{\odot}$, is shown in Fig. \ref{Fe_increase}. The white areas mark regions in the parameter space where stellar winds expel all the gas from the cloud as $R_{\textrm{SW}} = 0$. Therefore, in our scenario, these empty regions correspond to clusters that form only one stellar generation.  For the remaining regions of the parameter space, part of the initial leftover gas $M_{\textrm{gas}}$ has been retained ($M_{\textrm{ret}}>0$) and therefore may form a secondary stellar population with the metallicity enhancement as given in Figs. \ref{Fe_increase} and \ref{Fe_increase_b}. The stellar winds feedback increases with the star formation efficiency and with decreasing leftover gas mass (section \ref{wind_feedback}), which leads to larger gas expulsion regions as one move from the top to the bottom panels of Fig. \ref{Fe_increase}. Also note that the metallicity enhancement increases with the star formation efficiency and that rather small 1G star formation efficiencies are required to have $\Delta [\textrm{Fe/H}]\lesssim 0.1$ dex, the range observed between the first and the second stellar generations in galactic GCs.

Fig. \ref{Fe_increase_b} is similar to Fig. \ref{Fe_increase}. In this figure the upper and lower panels display the results for lower and and larger gas metallicities in the pre-stellar cloud: $Z/Z_{\odot}=10^{-2}$ and 1, respectively. The upper panels show that it is easier to increase the leftover gas metallicity in clouds with a lower initial gas metallicity (see \citealt{2019A&ARv..27....8G}). Therefore in these models the star formation efficiencies which leads to the requested $\Delta  \textrm{[Fe/H]}$ are lower than in models with Z $=10^{-1}$ Z$_{\odot}$. This implies that clusters with multiple stellar populations (MPs) are likely formed in clouds with low primordial gas metallicities. The bottom panels in Fig. \ref{Fe_increase_b} show that the opposite occurs for the case Z $=$ Z$_{\odot}$. Here, the 1G stars should form with a larger star-formation efficiency $\epsilon_{\textrm{1G}}$ to reach the same $\Delta \textrm{[Fe/H]}$ as in the previous, lower metallicity models. This enhances the ability of stellar winds to remove the leftover gas. Therefore it is likely that in such cases single population (SP) clusters are formed more often than in metal-poor clouds. 

The existence of a mass limit for clusters with MPs has been extensively debated in the literature \citep[e.g.][]{2011MNRAS.417..228C, 2013ApJ...778..186V, 2014ApJ...791L...4D, 2018MNRAS.477.4696M, 2019MNRAS.484.4718H}. However, the formation of stellar clusters with multiple populations is a multi-parameter problem. Indeed, the interplay among the proto-cluster cloud gas metallicity, core radius, mass and star-formation efficiency determines whether or not a cluster can form several stellar populations. For instance, in the model presented in the right bottom panel of Fig. \ref{Fe_increase}, a first stellar generation with mass $M_{1G}=10^{5}$ $M_{\odot}$ forms with an efficiency $\epsilon_{\textrm{1G}}=0.1$ from a cloud with Z $=10^{-1}$ Z$_{\odot}$ and core radius $R_{c}=1$ pc. In this case,  a second generation is allowed, as not all the leftover gas is removed out of the cloud by the 1G stellar winds. However, if $R_{c}=3$ pc, the secondary generation cannot be formed. For models with $\epsilon_{\textrm{1G}}=0.03$, a $M_{1G}=10^{5}$ $M_{\odot}$ cluster can form secondary generations whether its core radius is 1 or 3 pc. Clusters of the same mass with $\epsilon_{\textrm{1G}}=0.25$ but larger metallicity (Z $=$ Z$_{\odot}$), cannot form more than one generation unless it is very compact ($R_{c}<1$ pc, see the bottom right panel of Fig. \ref{Fe_increase_b}). 

In order to gain insight in the possible formation of multiple populations, Table \ref{tabla2} presents the output parameters of the models of Table \ref{tabla1}.  In this table, the total number of massive stars, which are expected to become supernovae, as well as the normalized super-wind and blowout radii, the number of trapped SNe, and the iron mass deposited by those SNe are presented in columns 2-6, respectively.  The 7th and the last column in Table \ref{tabla2} present the metallicity difference between the primordial cloud and the leftover gas contaminated by the 1G SNe, and the amount of leftover  gas available for the formation of 2G stars $M_{\textrm{ret}}$ (the mass of leftover gas within a sphere with radius $R_{\textrm{SW}}$). Note that in massive and compact low metallicity clouds (models 1-4), stellar winds manage to expel gas only from the outermost regions (see column 8 of Table \ref{tabla2}) for both core radii considered in this case ($R_{c}$=1.5 and 4 pc). Also, note that the metallicity enhancements are $\Delta \textrm{[Fe/H]}>0.1$ dex. Hence, a large fraction of the initial gas, enriched by the 1G stars, remains available for the 2G formation ($M_{\textrm{ret}}/M_{\textrm{gas}} \ge 0.95$, see also Fig. \ref{mass_retention}). Assuming that a second generation of stars form with mass $M_{\textrm{2G}}$ while 
exhausting all the enriched leftover gas $M_{\textrm{ret}}$ ($\epsilon_{2G}=1$), then we can estimate a lower limit for the fraction of 1G stars $f_{\textrm{1G}}=M_{\textrm{1G}}/(M_{\textrm{1G}}+M_{\textrm{2G}})$. For models 1-4, this fraction can be very small ($f_{\textrm{1G}}<0.1$). Similar conditions could potentially explain extreme observed cases such as $\omega$-Cen, where the enhancement of $\Delta \textrm{[Fe/H]}$ between the first and the second stellar generations is $\approx 0.3$ \citep{2008MNRAS.391..354R} and the observed fraction of 1G stars is $f_{\textrm{1G}}=0.086$ \citep{2017MNRAS.464.3636M}. 

For the intermediate metallicity $Z=10^{-1}$ Z$_{\odot}$, that correspond to models 5-8 in Tables \ref{tabla1}-\ref{tabla2}, the outcome is very sensitive on $\epsilon_{\textrm{1G}}$ and $R_{c}$. Indeed, models 5 and 7 (compact clusters) could form 2G stars such that $f_{\textrm{1G}}$ may be as small as 0.1. The fraction raises for the model 6 to $f_{\textrm{1G}}=0.21$ and note that the model 8 is a single population cluster ($f_{\textrm{1G}}=1$). So, if the gas metallicity is $Z=10^{-1}$ Z$_{\odot}$, the cluster with $M_{\textrm{1G}}= 2 \times 10^{5} M_{\odot}$ may form secondary populations such that $f_{\textrm{1G}}=$0.1 - 1, with the exact value determined by the core radius and the star-formation efficiencies. Finally, models 9-12 present cases with solar metallicity clouds. Note that in all but in case 9, the cluster would not host MPs stars. Hence, even for a given cluster mass ($M_{\textrm{1G}}= 2 \times 10^{5} M_{\odot}$), whether there could be MPs or not depend also on $R_{c}$, $\epsilon_{\textrm{1G}}$ and $Z$. This is in agreement with the observational data. For instance, the LMC cluster NGC 1783 and the SMC cluster NGC 419, whose masses are around $2 \times 10^{5} M_{\odot}$, are single populations clusters \citep{2017MNRAS.468.3150M, 2018ApJ...853..186Z}. However, multiple populations have been discovered for the LMC cluster NGC 1978 and the LMC cluster Hodge 6, which have comparable ages and masses (see \citealt{2018MNRAS.477.4696M,2019MNRAS.484.4718H}). More generally, our results are in agreement with the large scatter that there exist in the observed anti-correlation $M_{\textrm{cluster}}$-$f_{\textrm{1G}}$ \citep[e.g.][]{2018ARA&A..56...83B,2019A&ARv..27....8G}.

\begin{figure}
	\includegraphics[width=1\columnwidth]{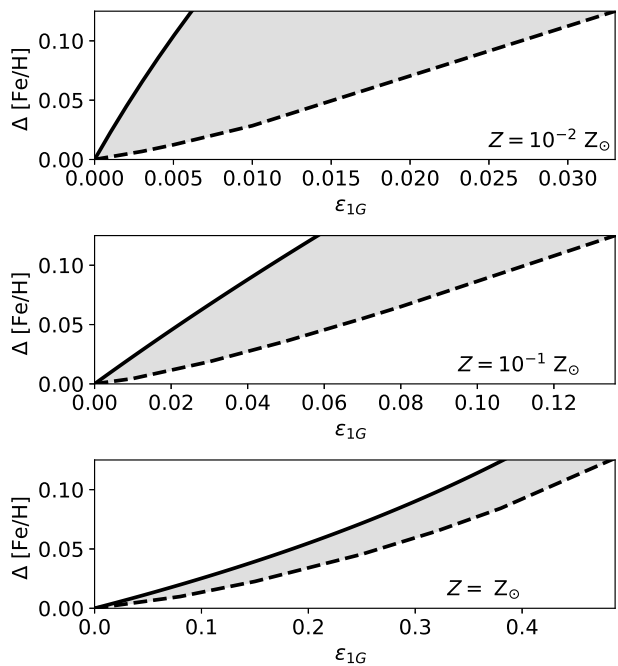}
    \caption{The iron enhancement $\Delta [\textrm{Fe/H}]$ as a function of
the star formation efficiency $\epsilon_{\textrm{1G}}$ of the first stellar generation
in massive, compact star-forming clouds. The panels consider the full range of
primordial cloud metallicities detected in Type I GCs. For a given
$\epsilon_{\textrm{1G}}$, $\Delta [\textrm{Fe/H}]$ depends both on $R_c$ and $M_{\textrm{1G}}$ as shown in Fig. \ref{Fe_increase}. However, for each $\epsilon_{\textrm{1G}}$ there exists a minimum value of $\Delta [\textrm{Fe/H}]$ which represents the case when blowout is most effective. The dashed lines
represent this lower limit to the metallicity enhancement. On the opposite case, the solid lines present the extreme case in which all the SNRs are trapped within the cloud and hence, mix their metals with the left over gas leading to large values of $\Delta [\textrm{Fe/H}]$. In such extreme case, which is an upper limit, $\Delta [\textrm{Fe/H}]$ does not depend on the other cloud parameters ($R_{c}$ and $M_{\textrm{gas}}$). It is clear then that all proto-cluster clouds should be located in the shaded region bounded by these critical lines.}
    \label{IRON_epsilon}
\end{figure}  

Fig. \ref{Fe_increase} shows that, for the range of values of $M_{\textrm{1G}}$ and $R_{c}$ here considered, the minimum feasible values of $\Delta\textrm{[Fe/H]}$ are determined by the star formation efficiency $\epsilon_{\textrm{1G}}$, with smaller $\Delta\textrm{[Fe/H]}$ obtained for decreasing values of $\epsilon_{\textrm{1G}}$. Fig. \ref{IRON_epsilon} presents the minimum values of $\Delta\textrm{[Fe/H]}$ (dashed lines) obtained upon the assumptions formulated in the present paper and compares it with the maximum possible value of $\Delta\textrm{[Fe/H]}$ (solid lines) calculated upon the assumption that all SNe were trapped within the star-forming cloud and their products were mixed with the gas left over after the 1G formation. Each panel applies to cases with a given primordial gas
metallicity. The upper limits (solid lines) are calculated by noting that in the case of complete retention of SN ejecta, equation (\ref{metallicity_difference}) is:
\begin{equation}\label{max_Delta}
\Delta [\textrm{Fe/H}]=\log \left(10^{[\textrm{Fe/H}]_{0}}+\frac{7}{13}\frac{\epsilon_{\textrm{1G}}}{1-\epsilon_{\textrm{1G}}} \right)-[\textrm{Fe/H}]_{0}.
\end{equation}
Equation (\ref{massive_stars_total}) has been used to derive this equation.

The shaded region between these two lines thus contains all values of $\Delta\textrm{[Fe/H]}$ allowed by the blowout model. Figures \ref{Fe_increase} and \ref{IRON_epsilon} show that star-forming clouds with a rather low 1G star formation efficiencies are required in order to obtain the metallicity enhancements observed between the first two stellar generations in Type I clusters. The allowed star formation efficiencies depend on the primordial metallicity. For example, massive ($10^6 -10^8$ M$_\odot$) compact clouds ($R_{c}\lesssim 3$ pc) undergoing a first burst of star formation with a low star formation efficiency ($\epsilon_{\textrm{1G}} \lesssim 0.027$) are required to match the properties of Type I globular clusters if $Z=10^{-2}$ Z$_{\odot}$. However, the maximum values of $\epsilon_{\textrm{1G}}$ are 0.11 and 0.42 for the larger metallicities $Z=10^{-1}$ Z$_{\odot}$ and Z$_{\odot}$, respectively. Unfortunately, our model do not allow one to fix the star formation efficiency of the second stellar generation, nor to predict the amount of low mass stars lost during the long evolution of globular clusters. However, if one knows the Fe content of the third and further stellar generations, one could define the star formation efficiency $\epsilon$ of all but the last stellar generation and get a handle also on the star formation track of Type II globular clusters. 
%---------------------------------------------------------------------
\section{SUMMARY AND CONCLUSIONS}\label{summarySection}
One of the central issues in the multiple populations problem is the uniform Fe metallicity in Type I globular clusters, as this implies that supernova ejecta from former stellar generations were not retained or captured within the gas the subsequent stellar generations formed from. We have taken here the point of view that SNe from stellar clusters do not produce a continuous wind, as usually assumed for the feedback from clusters in numerical calculations, but rather that they are independent events, hardly at all coincident in space and time, and able to form each their own remnant (a discussion of this assumption is given in Appendix \ref{Ap3}). We have followed the suggestion by \cite{tenorio2015supernovae} that supernova remnants may undergo blowout and through such events eject their metals and energy out of the proto-GC cloud, inhibiting or lowering the contamination of the leftover gas and thus of any future generation of stars. 

We have performed multiple simulations of the SNRs evolution in star-forming clouds with different mass, core radius, star formation efficiency and gas metallicity, by means of an updated version of the 3D thin shell approximation code (see \citealt{2019MNRAS.488..978J}). The code assumes a Gaussian density distribution in the parental cloud and in the emerging cluster and accounts for the mass and kinetic energy of the ejecta of each exploding star. 

In all cases we have found the radius $R_{\textrm{SW}}$ which marks the inner location from which a cluster wind could develop and the blowout radius $R_{\textrm{blow}}$, or minimum distance away from the cluster center where a $10^{51}$ erg SN explosion would lead to blowout. We have shown that $R_{\textrm{blow}}$ depends exclusively on the gas concentration, which scales directly with the mass of the leftover gas and inversely with the cloud core radius (see equation \ref{blow_function}). SNRs that exploded within this radius are pressure confined and thus should mix their products with the leftover gas. In order to study which initial conditions lead to the observed small iron spreads in Type I GCs ($\lesssim 0.1$ dex; see \citealt{2009A&A...508..695C}, \citealt{2020IAUS..351..251M}), the iron mass returned by each such SN was added to the leftover gas.

The results of the calculations are presented in Figures \ref{Fe_increase} and \ref{Fe_increase_b}
while Tables \ref{tabla1} and \ref{tabla2} present several examples. We have considered star formation events at high densities and thus in massive and compact clouds. This favours the SN blowout, the retention or survival of a large cloud mass after being confronted with the winds from massive stars and also favours the sufficient dilution of the trapped SN ejecta to warrant a low metallicity enhancement of the second stellar generation. Our results can be summarized as:
\begin{enumerate}
\item The ability of stellar winds to clear out the gas left over from star formation depends on the cloud mass and core radius. Massive ($10^{6}$-$10^{8}$ M$_{\odot}$) and compact clouds ($R_{c} \lesssim 3$ pc) can retain large fractions of their leftover gas within their central zones \citep{Silich2017,Silich2018,Silich2020}. Such conditions may not be typical for present day Milky Way clouds, but are expected for large pressure environments in merging systems and star-forming galaxies at redshifts $z > 2$ \citep[e.g.][]{2015ApJ...806...35J, 2014CQGra..31x4006K,2015MNRAS.454.1658K}. In models with larger core radius the individual stellar winds merge to form a global star cluster wind and expel the leftover gas into the ambient ISM. In such cases, further episodes of star formation may only be possible through gas accretion at later times \citep[e.g.][]{2015ApJ...814L..14C,2019MNRAS.489.3269C,2016MNRAS.461.4088D}. Note however that, to end up with a type I GC, the metallicity of the accreted cloud ought to be smaller or very similar to that of the cloud that gave origin to the first stellar generation.
\item The star formation efficiency and the initial gas metallicity also influence the gas expulsion by stellar winds. Gas retention is more likely for small efficiencies and metal-poor clouds. 

\item Constraining our models such that the metallicity spread between consecutive stellar generations is similar to those observed in Type I GCs ($\Delta \textrm{[Fe/H]}  \sim$  0.1 dex), we have determined an upper limit for the star formation efficiency of the first stellar generation ($\epsilon_{\textrm{1G}}$). Furthermore, the upper limit for $\epsilon_{\textrm{1G}}$ grows with the metallicity of the progenitor cloud ($\epsilon_{\textrm{1G}} \lesssim 0.027, 0.11$ and 0.42 for $Z/Z_{\odot}=10^{-2}, 10^{-1}$ and 1, respectively). This is a consequence both of the SN blowout and of the fact that much more enriched material is required to obtain such abundance spreads ($\Delta \textrm{[Fe/H]}  \sim$  0.1 dex) in metal-rich clusters.  
\item Finally, our results indicate that the 1G mass is not the only parameter that constraint the formation of multiple populations. The core radius, star formation efficiency and gas metallicity of the progenitor cloud are also relevant parameters. Clusters with equal 1G stellar mass $M_{\textrm{1G}}$ can form different fractions of enriched (2G) stars. 
\end{enumerate}

Given the minute amount of matter lost from the parental cloud after the blowout of each SN, one should expect several stellar generations in a GC. This is to be expected for as long as a global cluster wind is inhibited. Once the steering and mixing caused by SN explosions stops, a further collapse proceeds causing the formation of a new stellar generation. The amount of gas in the massive star-forming cloud decreases after being transformed into stars in every star formation event and due to partial expulsion by stellar winds. This is the main loss of gas, although some of it comes back as stellar winds and trapped SNe ejecta. As we have shown here, if these events occur with a low star formation efficiency, perhaps due to the low metal abundance and small cooling, the resultant generations will present a small, hardly noticeable, $\Delta \textrm{[Fe/H]}$, smaller than 0.1 dex. Larger star formation efficiencies would lead to larger values of $\Delta \textrm{[Fe/H]}$ leading to very different metallicities between  consecutive stellar  generations, as in Type II globular clusters, subject of our forthcoming communication.

\section*{Acknowledgements}
We thank our anonymous referee for timely and constructive comments that improved the presentation of our results. We would like to thank Sergio Mart\'inez-Gonz\'alez, for helpful discussions on the numerical calculations. This study was supported by CONACYT-M\'exico research grant A1-S-28458. SJ acknowledge the support by CONACYT-M\'exico (scholarship registration number 613136). The authors thankfully acknowledge the support provided by the Laboratorio Nacional de Superc\'omputo del Sureste de M\'exico, CONACYT member of the network of national laboratories.
%%%%%%%%%%%%%%%%%%%%%%%%%%%%%%%%%%%%%%%%%%%%%%%%%%



%\DeclareRobustCommand{\DE}[3]{#3}

\section*{DATA AVAILABILITY STATEMENT}
The data underlying this article will be shared on reasonable request
to the corresponding author.

\bibliographystyle{mnras}

\bibliography{ref}

\providecommand{\noopsort}[1]{}
\begin{thebibliography}{}
\makeatletter
\relax
\def\mn@urlcharsother{\let\do\@makeother \do\$\do\&\do\#\do\^\do\_\do\%\do\~}
\def\mn@doi{\begingroup\mn@urlcharsother \@ifnextchar [ {\mn@doi@}
  {\mn@doi@[]}}
\def\mn@doi@[#1]#2{\def\@tempa{#1}\ifx\@tempa\@empty \href
  {http://dx.doi.org/#2} {doi:#2}\else \href {http://dx.doi.org/#2} {#1}\fi
  \endgroup}
\def\mn@eprint#1#2{\mn@eprint@#1:#2::\@nil}
\def\mn@eprint@arXiv#1{\href {http://arxiv.org/abs/#1} {{\tt arXiv:#1}}}
\def\mn@eprint@dblp#1{\href {http://dblp.uni-trier.de/rec/bibtex/#1.xml}
  {dblp:#1}}
\def\mn@eprint@#1:#2:#3:#4\@nil{\def\@tempa {#1}\def\@tempb {#2}\def\@tempc
  {#3}\ifx \@tempc \@empty \let \@tempc \@tempb \let \@tempb \@tempa \fi \ifx
  \@tempb \@empty \def\@tempb {arXiv}\fi \@ifundefined
  {mn@eprint@\@tempb}{\@tempb:\@tempc}{\expandafter \expandafter \csname
  mn@eprint@\@tempb\endcsname \expandafter{\@tempc}}}

\bibitem[\protect\citeauthoryear{{Bastian} \& {Lardo}}{{Bastian} \&
  {Lardo}}{2018}]{2018ARA&A..56...83B}
{Bastian} N.,  {Lardo} C.,  2018, \mn@doi [\araa]
  {10.1146/annurev-astro-081817-051839}, \href
  {https://ui.adsabs.harvard.edu/abs/2018ARA%26A..56...83B} {56, 83}

\bibitem[\protect\citeauthoryear{{Bastian}, {Lamers}, {de Mink}, {Longmore},
  {Goodwin}  \& {Gieles}}{{Bastian} et~al.}{2013}]{2013MNRAS.436.2398B}
{Bastian} N.,  {Lamers} H.~J.~G.~L.~M.,  {de Mink} S.~E.,  {Longmore} S.~N.,
  {Goodwin} S.~P.,   {Gieles} M.,  2013, \mn@doi [\mnras]
  {10.1093/mnras/stt1745}, \href
  {https://ui.adsabs.harvard.edu/abs/2013MNRAS.436.2398B} {436, 2398}

\bibitem[\protect\citeauthoryear{{Baumgartner} \&
  {Breitschwerdt}}{{Baumgartner} \&
  {Breitschwerdt}}{2013}]{2013A&A...557A.140B}
{Baumgartner} V.,  {Breitschwerdt} D.,  2013, \mn@doi [\aap]
  {10.1051/0004-6361/201321261}, \href
  {https://ui.adsabs.harvard.edu/abs/2013A%26A...557A.140B} {557, A140}

\bibitem[\protect\citeauthoryear{{Bedin}, {Piotto}, {Anderson}, {Cassisi},
  {King}, {Momany}  \& {Carraro}}{{Bedin} et~al.}{2004}]{2004ApJ...605L.125B}
{Bedin} L.~R.,  {Piotto} G.,  {Anderson} J.,  {Cassisi} S.,  {King} I.~R.,
  {Momany} Y.,   {Carraro} G.,  2004, \mn@doi [\apjl] {10.1086/420847}, \href
  {https://ui.adsabs.harvard.edu/abs/2004ApJ...605L.125B} {605, L125}

\bibitem[\protect\citeauthoryear{{Bisnovatyi-Kogan} \&
  {Silich}}{{Bisnovatyi-Kogan} \& {Silich}}{1995}]{bisno1}
{Bisnovatyi-Kogan} G.~S.,  {Silich} S.~A.,  1995, \mn@doi [Reviews of Modern
  Physics] {10.1103/RevModPhys.67.661}, \href
  {http://adsabs.harvard.edu/abs/1995RvMP...67..661B} {67, 661}

\bibitem[\protect\citeauthoryear{{Blondin}, {Wright}, {Borkowski}  \&
  {Reynolds}}{{Blondin} et~al.}{1998}]{Blondin1998}
{Blondin} J.~M.,  {Wright} E.~B.,  {Borkowski} K.~J.,   {Reynolds} S.~P.,
  1998, \mn@doi [\apj] {10.1086/305708}, \href
  {http://adsabs.harvard.edu/abs/1998ApJ...500..342B} {500, 342}

\bibitem[\protect\citeauthoryear{{Caloi} \& {D'Antona}}{{Caloi} \&
  {D'Antona}}{2011}]{2011MNRAS.417..228C}
{Caloi} V.,  {D'Antona} F.,  2011, \mn@doi [\mnras]
  {10.1111/j.1365-2966.2011.19166.x}, \href
  {https://ui.adsabs.harvard.edu/abs/2011MNRAS.417..228C} {417, 228}

\bibitem[\protect\citeauthoryear{{Calura}, {Few}, {Romano}  \&
  {D'Ercole}}{{Calura} et~al.}{2015}]{2015ApJ...814L..14C}
{Calura} F.,  {Few} C.~G.,  {Romano} D.,   {D'Ercole} A.,  2015, \mn@doi
  [\apjl] {10.1088/2041-8205/814/1/L14}, \href
  {https://ui.adsabs.harvard.edu/abs/2015ApJ...814L..14C} {814, L14}

\bibitem[\protect\citeauthoryear{{Calura}, {D'Ercole}, {Vesperini}, {Vanzella}
  \& {Sollima}}{{Calura} et~al.}{2019}]{2019MNRAS.489.3269C}
{Calura} F.,  {D'Ercole} A.,  {Vesperini} E.,  {Vanzella} E.,   {Sollima} A.,
  2019, \mn@doi [\mnras] {10.1093/mnras/stz2055}, \href
  {https://ui.adsabs.harvard.edu/abs/2019MNRAS.489.3269C} {489, 3269}

\bibitem[\protect\citeauthoryear{{Carretta}}{{Carretta}}{2019}]{2019A&A...624A..24C}
{Carretta} E.,  2019, \mn@doi [\aap] {10.1051/0004-6361/201935110}, \href
  {https://ui.adsabs.harvard.edu/abs/2019A&A...624A..24C} {624, A24}

\bibitem[\protect\citeauthoryear{{Carretta}, {Bragaglia}, {Gratton}, {D'Orazi}
  \& {Lucatello}}{{Carretta} et~al.}{2009}]{2009A&A...508..695C}
{Carretta} E.,  {Bragaglia} A.,  {Gratton} R.,  {D'Orazi} V.,   {Lucatello} S.,
   2009, \mn@doi [\aap] {10.1051/0004-6361/200913003}, \href
  {https://ui.adsabs.harvard.edu/abs/2009A&A...508..695C} {508, 695}

\bibitem[\protect\citeauthoryear{{Carretta} et~al.,}{{Carretta}
  et~al.}{2015}]{2015A&A...578A.116C}
{Carretta} E.,  et~al., 2015, \mn@doi [\aap] {10.1051/0004-6361/201525951},
  \href {https://ui.adsabs.harvard.edu/abs/2015A%26A...578A.116C} {578, A116}

\bibitem[\protect\citeauthoryear{{Chevalier}}{{Chevalier}}{1977}]{1977Chevalier}
{Chevalier} R.~A.,  1977, \mn@doi [\araa]
  {10.1146/annurev.aa.15.090177.001135}, \href
  {http://adsabs.harvard.edu/abs/1977ARA%26A..15..175C} {15, 175}

\bibitem[\protect\citeauthoryear{{Chevalier}}{{Chevalier}}{1982}]{chevalier1982self}
{Chevalier} R.~A.,  1982, \mn@doi [\apj] {10.1086/160126}, \href
  {http://adsabs.harvard.edu/abs/1982ApJ...258..790C} {258, 790}

\bibitem[\protect\citeauthoryear{{Chevalier}}{{Chevalier}}{1984}]{chevalier1984interaction}
{Chevalier} R.~A.,  1984, \mn@doi [Annals of the New York Academy of Sciences]
  {10.1111/j.1749-6632.1984.tb23355.x}, \href
  {http://adsabs.harvard.edu/abs/1984NYASA.422..215C} {422, 215}

\bibitem[\protect\citeauthoryear{{Cioffi}, {McKee}  \& {Bertschinger}}{{Cioffi}
  et~al.}{1988}]{cioffi1988dynamics}
{Cioffi} D.~F.,  {McKee} C.~F.,   {Bertschinger} E.,  1988, \mn@doi [\apj]
  {10.1086/166834}, \href {http://adsabs.harvard.edu/abs/1988ApJ...334..252C}
  {334, 252}

\bibitem[\protect\citeauthoryear{{D'Antona} \& {Caloi}}{{D'Antona} \&
  {Caloi}}{2004}]{2004ApJ...611..871D}
{D'Antona} F.,  {Caloi} V.,  2004, \mn@doi [\apj] {10.1086/422334}, \href
  {https://ui.adsabs.harvard.edu/abs/2004ApJ...611..871D} {611, 871}

\bibitem[\protect\citeauthoryear{{D'Antona}, {Caloi}, {Montalb{\'a}n},
  {Ventura}  \& {Gratton}}{{D'Antona} et~al.}{2002}]{2002A&A...395...69D}
{D'Antona} F.,  {Caloi} V.,  {Montalb{\'a}n} J.,  {Ventura} P.,   {Gratton} R.,
   2002, \mn@doi [\aap] {10.1051/0004-6361:20021220}, \href
  {https://ui.adsabs.harvard.edu/abs/2002A%26A...395...69D} {395, 69}

\bibitem[\protect\citeauthoryear{{D'Antona}, {Vesperini}, {D'Ercole},
  {Ventura}, {Milone}, {Marino}  \& {Tailo}}{{D'Antona}
  et~al.}{2016}]{2016MNRAS.458.2122D}
{D'Antona} F.,  {Vesperini} E.,  {D'Ercole} A.,  {Ventura} P.,  {Milone} A.~P.,
   {Marino} A.~F.,   {Tailo} M.,  2016, \mn@doi [\mnras]
  {10.1093/mnras/stw387}, \href
  {https://ui.adsabs.harvard.edu/abs/2016MNRAS.458.2122D} {458, 2122}

\bibitem[\protect\citeauthoryear{{D'Ercole}, {D'Antona}, {Ventura}, {Vesperini}
   \& {McMillan}}{{D'Ercole} et~al.}{2010}]{2010MNRAS.407..854D}
{D'Ercole} A.,  {D'Antona} F.,  {Ventura} P.,  {Vesperini} E.,   {McMillan}
  S.~L.~W.,  2010, \mn@doi [\mnras] {10.1111/j.1365-2966.2010.16996.x}, \href
  {https://ui.adsabs.harvard.edu/abs/2010MNRAS.407..854D} {407, 854}

\bibitem[\protect\citeauthoryear{{D'Ercole}, {D'Antona}  \&
  {Vesperini}}{{D'Ercole} et~al.}{2016}]{2016MNRAS.461.4088D}
{D'Ercole} A.,  {D'Antona} F.,   {Vesperini} E.,  2016, \mn@doi [\mnras]
  {10.1093/mnras/stw1583}, \href
  {https://ui.adsabs.harvard.edu/abs/2016MNRAS.461.4088D} {461, 4088}

\bibitem[\protect\citeauthoryear{{Da Costa}, {Held}, {Saviane}  \&
  {Gullieuszik}}{{Da Costa} et~al.}{2009}]{2009ApJ...705.1481D}
{Da Costa} G.~S.,  {Held} E.~V.,  {Saviane} I.,   {Gullieuszik} M.,  2009,
  \mn@doi [\apj] {10.1088/0004-637X/705/2/1481}, \href
  {https://ui.adsabs.harvard.edu/abs/2009ApJ...705.1481D} {705, 1481}

\bibitem[\protect\citeauthoryear{{Dale}, {Ercolano}  \& {Bonnell}}{{Dale}
  et~al.}{2015}]{2015MNRAS.451..987D}
{Dale} J.~E.,  {Ercolano} B.,   {Bonnell} I.~A.,  2015, \mn@doi [\mnras]
  {10.1093/mnras/stv913}, \href
  {https://ui.adsabs.harvard.edu/abs/2015MNRAS.451..987D} {451, 987}

\bibitem[\protect\citeauthoryear{{Dalessandro} et~al.,}{{Dalessandro}
  et~al.}{2014}]{2014ApJ...791L...4D}
{Dalessandro} E.,  et~al., 2014, \mn@doi [\apjl] {10.1088/2041-8205/791/1/L4},
  \href {https://ui.adsabs.harvard.edu/abs/2014ApJ...791L...4D} {791, L4}

\bibitem[\protect\citeauthoryear{{Dalessandro}, {Lapenna}, {Mucciarelli},
  {Origlia}, {Ferraro}  \& {Lanzoni}}{{Dalessandro}
  et~al.}{2016}]{2016ApJ...829...77D}
{Dalessandro} E.,  {Lapenna} E.,  {Mucciarelli} A.,  {Origlia} L.,  {Ferraro}
  F.~R.,   {Lanzoni} B.,  2016, \mn@doi [\apj] {10.3847/0004-637X/829/2/77},
  \href {https://ui.adsabs.harvard.edu/abs/2016ApJ...829...77D} {829, 77}

\bibitem[\protect\citeauthoryear{{Decressin}, {Meynet}, {Charbonnel},
  {Prantzos}  \& {Ekstr{\"o}m}}{{Decressin}
  et~al.}{2007a}]{2007A&A...464.1029D}
{Decressin} T.,  {Meynet} G.,  {Charbonnel} C.,  {Prantzos} N.,   {Ekstr{\"o}m}
  S.,  2007a, \mn@doi [\aap] {10.1051/0004-6361:20066013}, \href
  {https://ui.adsabs.harvard.edu/abs/2007A%26A...464.1029D} {464, 1029}

\bibitem[\protect\citeauthoryear{{Decressin}, {Charbonnel}  \&
  {Meynet}}{{Decressin} et~al.}{2007b}]{2007A&A...475..859D}
{Decressin} T.,  {Charbonnel} C.,   {Meynet} G.,  2007b, \mn@doi [\aap]
  {10.1051/0004-6361:20078425}, \href
  {https://ui.adsabs.harvard.edu/abs/2007A%26A...475..859D} {475, 859}

\bibitem[\protect\citeauthoryear{{Denissenkov} \& {Hartwick}}{{Denissenkov} \&
  {Hartwick}}{2014}]{2014MNRAS.437L..21D}
{Denissenkov} P.~A.,  {Hartwick} F.~D.~A.,  2014, \mn@doi [\mnras]
  {10.1093/mnrasl/slt133}, \href
  {https://ui.adsabs.harvard.edu/abs/2014MNRAS.437L..21D} {437, L21}

\bibitem[\protect\citeauthoryear{{Draine} \& {McKee}}{{Draine} \&
  {McKee}}{1993}]{draine1993theory}
{Draine} B.~T.,  {McKee} C.~F.,  1993, \mn@doi [\araa]
  {10.1146/annurev.aa.31.090193.002105}, \href
  {http://adsabs.harvard.edu/abs/1993ARA%26A..31..373D} {31, 373}

\bibitem[\protect\citeauthoryear{{Farias}, {Fellhauer}, {Smith},
  {Dom{\'\i}nguez}  \& {Dabringhausen}}{{Farias}
  et~al.}{2018}]{2018MNRAS.476.5341F}
{Farias} J.~P.,  {Fellhauer} M.,  {Smith} R.,  {Dom{\'\i}nguez} R.,
  {Dabringhausen} J.,  2018, \mn@doi [\mnras] {10.1093/mnras/sty597}, \href
  {https://ui.adsabs.harvard.edu/abs/2018MNRAS.476.5341F} {476, 5341}

\bibitem[\protect\citeauthoryear{{Fern{\'a}ndez-Trincado}
  et~al.,}{{Fern{\'a}ndez-Trincado} et~al.}{2021}]{2021arXiv210207785F}
{Fern{\'a}ndez-Trincado} J.~G.,  et~al., 2021, arXiv e-prints, \href
  {https://ui.adsabs.harvard.edu/abs/2021arXiv210207785F} {p. arXiv:2102.07785}

\bibitem[\protect\citeauthoryear{{Gieles} et~al.,}{{Gieles}
  et~al.}{2018}]{2018MNRAS.478.2461G}
{Gieles} M.,  et~al., 2018, \mn@doi [\mnras] {10.1093/mnras/sty1059}, \href
  {https://ui.adsabs.harvard.edu/abs/2018MNRAS.478.2461G} {478, 2461}

\bibitem[\protect\citeauthoryear{{Gilligan} et~al.,}{{Gilligan}
  et~al.}{2019}]{2019MNRAS.486.5581G}
{Gilligan} C.~K.,  et~al., 2019, \mn@doi [\mnras] {10.1093/mnras/stz1174},
  \href {https://ui.adsabs.harvard.edu/abs/2019MNRAS.486.5581G} {486, 5581}

\bibitem[\protect\citeauthoryear{{Goodwin} \& {Bastian}}{{Goodwin} \&
  {Bastian}}{2006}]{2006MNRAS.373..752G}
{Goodwin} S.~P.,  {Bastian} N.,  2006, \mn@doi [\mnras]
  {10.1111/j.1365-2966.2006.11078.x}, \href
  {https://ui.adsabs.harvard.edu/abs/2006MNRAS.373..752G} {373, 752}

\bibitem[\protect\citeauthoryear{{Gratton}, {Villanova}, {Lucatello},
  {Sollima}, {Geisler}, {Carretta}, {Cassisi}  \& {Bragaglia}}{{Gratton}
  et~al.}{2012}]{2012A&A...544A..12G}
{Gratton} R.~G.,  {Villanova} S.,  {Lucatello} S.,  {Sollima} A.,  {Geisler}
  D.,  {Carretta} E.,  {Cassisi} S.,   {Bragaglia} A.,  2012, \mn@doi [\aap]
  {10.1051/0004-6361/201219276}, \href
  {https://ui.adsabs.harvard.edu/abs/2012A%26A...544A..12G} {544, A12}

\bibitem[\protect\citeauthoryear{{Gratton}, {Bragaglia}, {Carretta}, {D'Orazi},
  {Lucatello}  \& {Sollima}}{{Gratton} et~al.}{2019}]{2019A&ARv..27....8G}
{Gratton} R.,  {Bragaglia} A.,  {Carretta} E.,  {D'Orazi} V.,  {Lucatello} S.,
   {Sollima} A.,  2019, \mn@doi [\aapr] {10.1007/s00159-019-0119-3}, \href
  {https://ui.adsabs.harvard.edu/abs/2019A&ARv..27....8G} {27, 8}

\bibitem[\protect\citeauthoryear{{Hamilton} \& {Sarazin}}{{Hamilton} \&
  {Sarazin}}{1984}]{hamilton1984new}
{Hamilton} A.~J.~S.,  {Sarazin} C.~L.,  1984, \mn@doi [\apj] {10.1086/162145},
  \href {http://adsabs.harvard.edu/abs/1984ApJ...281..682H} {281, 682}

\bibitem[\protect\citeauthoryear{{Hamuy}}{{Hamuy}}{2003}]{2003ApJ...582..905H}
{Hamuy} M.,  2003, \mn@doi [\apj] {10.1086/344689}, \href
  {https://ui.adsabs.harvard.edu/abs/2003ApJ...582..905H} {582, 905}

\bibitem[\protect\citeauthoryear{{Hollyhead} et~al.,}{{Hollyhead}
  et~al.}{2019}]{2019MNRAS.484.4718H}
{Hollyhead} K.,  et~al., 2019, \mn@doi [\mnras] {10.1093/mnras/stz317}, \href
  {https://ui.adsabs.harvard.edu/abs/2019MNRAS.484.4718H} {484, 4718}

\bibitem[\protect\citeauthoryear{{Jim{\'e}nez}, {Tenorio-Tagle}  \&
  {Silich}}{{Jim{\'e}nez} et~al.}{2019}]{2019MNRAS.488..978J}
{Jim{\'e}nez} S.,  {Tenorio-Tagle} G.,   {Silich} S.,  2019, \mn@doi [\mnras]
  {10.1093/mnras/stz1749}, \href
  {https://ui.adsabs.harvard.edu/abs/2019MNRAS.488..978J} {488, 978}

\bibitem[\protect\citeauthoryear{{Johnson}, {Rich}, {Pilachowski}, {Caldwell},
  {Mateo}, {Bailey}  \& {Crane}}{{Johnson} et~al.}{2015a}]{2015AJ....150...63J}
{Johnson} C.~I.,  {Rich} R.~M.,  {Pilachowski} C.~A.,  {Caldwell} N.,  {Mateo}
  M.,  {Bailey} John~I. I.,   {Crane} J.~D.,  2015a, \mn@doi [\aj]
  {10.1088/0004-6256/150/2/63}, \href
  {https://ui.adsabs.harvard.edu/abs/2015AJ....150...63J} {150, 63}

\bibitem[\protect\citeauthoryear{{Johnson}, {Leroy}, {Indebetouw}, {Brogan},
  {Whitmore}, {Hibbard}, {Sheth}  \& {Evans}}{{Johnson}
  et~al.}{2015b}]{2015ApJ...806...35J}
{Johnson} K.~E.,  {Leroy} A.~K.,  {Indebetouw} R.,  {Brogan} C.~L.,  {Whitmore}
  B.~C.,  {Hibbard} J.,  {Sheth} K.,   {Evans} A.~S.,  2015b, \mn@doi [\apj]
  {10.1088/0004-637X/806/1/35}, \href
  {https://ui.adsabs.harvard.edu/abs/2015ApJ...806...35J} {806, 35}

\bibitem[\protect\citeauthoryear{{Johnson}, {Caldwell}, {Rich}, {Mateo},
  {Bailey}, {Clarkson}, {Olszewski}  \& {Walker}}{{Johnson}
  et~al.}{2017}]{2017ApJ...836..168J}
{Johnson} C.~I.,  {Caldwell} N.,  {Rich} R.~M.,  {Mateo} M.,  {Bailey} John~I.
  I.,  {Clarkson} W.~I.,  {Olszewski} E.~W.,   {Walker} M.~G.,  2017, \mn@doi
  [\apj] {10.3847/1538-4357/836/2/168}, \href
  {https://ui.adsabs.harvard.edu/abs/2017ApJ...836..168J} {836, 168}

\bibitem[\protect\citeauthoryear{{Kim} \& {Ostriker}}{{Kim} \&
  {Ostriker}}{2015}]{KimOstriker}
{Kim} C.-G.,  {Ostriker} E.~C.,  2015, \mn@doi [\apj]
  {10.1088/0004-637X/802/2/99}, \href
  {http://adsabs.harvard.edu/abs/2015ApJ...802...99K} {802, 99}

\bibitem[\protect\citeauthoryear{{Koo} \& {McKee}}{{Koo} \&
  {McKee}}{1992}]{1992ApJ...388...93K}
{Koo} B.-C.,  {McKee} C.~F.,  1992, \mn@doi [\apj] {10.1086/171132}, \href
  {https://ui.adsabs.harvard.edu/abs/1992ApJ...388...93K} {388, 93}

\bibitem[\protect\citeauthoryear{{Krause}, {Charbonnel}, {Decressin}, {Meynet}
  \& {Prantzos}}{{Krause} et~al.}{2013}]{krause2013superbubble}
{Krause} M.,  {Charbonnel} C.,  {Decressin} T.,  {Meynet} G.,   {Prantzos} N.,
  2013, \mn@doi [\aap] {10.1051/0004-6361/201220694}, \href
  {http://adsabs.harvard.edu/abs/2013A%26A...552A.121K} {552, A121}

\bibitem[\protect\citeauthoryear{{Krause} et~al.,}{{Krause}
  et~al.}{2020}]{2020SSRv..216...64K}
{Krause} M. G.~H.,  et~al., 2020, \mn@doi [\ssr] {10.1007/s11214-020-00689-4},
  \href {https://ui.adsabs.harvard.edu/abs/2020SSRv..216...64K} {216, 64}

\bibitem[\protect\citeauthoryear{{Kruijssen}}{{Kruijssen}}{2014}]{2014CQGra..31x4006K}
{Kruijssen} J.~M.~D.,  2014, \mn@doi [Classical and Quantum Gravity]
  {10.1088/0264-9381/31/24/244006}, \href
  {https://ui.adsabs.harvard.edu/abs/2014CQGra..31x4006K} {31, 244006}

\bibitem[\protect\citeauthoryear{{Kruijssen}}{{Kruijssen}}{2015}]{2015MNRAS.454.1658K}
{Kruijssen} J.~M.~D.,  2015, \mn@doi [\mnras] {10.1093/mnras/stv2026}, \href
  {https://ui.adsabs.harvard.edu/abs/2015MNRAS.454.1658K} {454, 1658}

\bibitem[\protect\citeauthoryear{{Larsen}, {Brodie}, {Grundahl}  \&
  {Strader}}{{Larsen} et~al.}{2014}]{2014ApJ...797...15L}
{Larsen} S.~S.,  {Brodie} J.~P.,  {Grundahl} F.,   {Strader} J.,  2014, \mn@doi
  [\apj] {10.1088/0004-637X/797/1/15}, \href
  {https://ui.adsabs.harvard.edu/abs/2014ApJ...797...15L} {797, 15}

\bibitem[\protect\citeauthoryear{{Lee}, {Joo}, {Sohn}, {Rey}, {Lee}  \&
  {Walker}}{{Lee} et~al.}{1999}]{1999Natur.402...55L}
{Lee} Y.~W.,  {Joo} J.~M.,  {Sohn} Y.~J.,  {Rey} S.~C.,  {Lee} H.~C.,
  {Walker} A.~R.,  1999, \mn@doi [\nat] {10.1038/46985}, \href
  {https://ui.adsabs.harvard.edu/abs/1999Natur.402...55L} {402, 55}

\bibitem[\protect\citeauthoryear{{Leitherer} et~al.,}{{Leitherer}
  et~al.}{1999}]{1999ApJS..123....3L}
{Leitherer} C.,  et~al., 1999, \mn@doi [\apjs] {10.1086/313233}, \href
  {https://ui.adsabs.harvard.edu/abs/1999ApJS..123....3L} {123, 3}

\bibitem[\protect\citeauthoryear{{Li}, {Ostriker}, {Cen}, {Bryan}  \&
  {Naab}}{{Li} et~al.}{2015}]{LiOstriker}
{Li} M.,  {Ostriker} J.~P.,  {Cen} R.,  {Bryan} G.~L.,   {Naab} T.,  2015,
  \mn@doi [\apj] {10.1088/0004-637X/814/1/4}, \href
  {http://adsabs.harvard.edu/abs/2015ApJ...814....4L} {814, 4}

\bibitem[\protect\citeauthoryear{{Lochhaas} \& {Thompson}}{{Lochhaas} \&
  {Thompson}}{2017}]{2017MNRAS.470..977L}
{Lochhaas} C.,  {Thompson} T.~A.,  2017, \mn@doi [\mnras]
  {10.1093/mnras/stx1289}, \href
  {https://ui.adsabs.harvard.edu/abs/2017MNRAS.470..977L} {470, 977}

\bibitem[\protect\citeauthoryear{{Mac Low}, {McCray}  \& {Norman}}{{Mac Low}
  et~al.}{1989}]{1989ApJ...337..141M}
{Mac Low} M.-M.,  {McCray} R.,   {Norman} M.~L.,  1989, \mn@doi [\apj]
  {10.1086/167094}, \href
  {https://ui.adsabs.harvard.edu/abs/1989ApJ...337..141M} {337, 141}

\bibitem[\protect\citeauthoryear{{Marino} et~al.,}{{Marino}
  et~al.}{2015}]{2015MNRAS.450..815M}
{Marino} A.~F.,  et~al., 2015, \mn@doi [\mnras] {10.1093/mnras/stv420}, \href
  {https://ui.adsabs.harvard.edu/abs/2015MNRAS.450..815M} {450, 815}

\bibitem[\protect\citeauthoryear{{Marino} et~al.,}{{Marino}
  et~al.}{2019}]{2019MNRAS.487.3815M}
{Marino} A.~F.,  et~al., 2019, \mn@doi [\mnras] {10.1093/mnras/stz1415}, \href
  {https://ui.adsabs.harvard.edu/abs/2019MNRAS.487.3815M} {487, 3815}

\bibitem[\protect\citeauthoryear{{Martocchia} et~al.,}{{Martocchia}
  et~al.}{2017}]{2017MNRAS.468.3150M}
{Martocchia} S.,  et~al., 2017, \mn@doi [\mnras] {10.1093/mnras/stx660}, \href
  {https://ui.adsabs.harvard.edu/abs/2017MNRAS.468.3150M} {468, 3150}

\bibitem[\protect\citeauthoryear{{Martocchia} et~al.,}{{Martocchia}
  et~al.}{2018}]{2018MNRAS.477.4696M}
{Martocchia} S.,  et~al., 2018, \mn@doi [\mnras] {10.1093/mnras/sty916}, \href
  {https://ui.adsabs.harvard.edu/abs/2018MNRAS.477.4696M} {477, 4696}

\bibitem[\protect\citeauthoryear{{Martocchia} et~al.,}{{Martocchia}
  et~al.}{2019}]{2019MNRAS.487.5324M}
{Martocchia} S.,  et~al., 2019, \mn@doi [\mnras] {10.1093/mnras/stz1596}, \href
  {https://ui.adsabs.harvard.edu/abs/2019MNRAS.487.5324M} {487, 5324}

\bibitem[\protect\citeauthoryear{{Milone}}{{Milone}}{2020}]{2020IAUS..351..251M}
{Milone} A.~P.,  2020, in {Bragaglia} A.,  {Davies} M.,  {Sills} A.,
  {Vesperini} E.,  eds,  IAU Symposium Vol. 351, IAU Symposium. pp 251--260
  (\mn@eprint {arXiv} {1908.11703}), \mn@doi{10.1017/S1743921319010044}

\bibitem[\protect\citeauthoryear{{Milone} et~al.,}{{Milone}
  et~al.}{2015a}]{2015MNRAS.447..927M}
{Milone} A.~P.,  et~al., 2015a, \mn@doi [\mnras] {10.1093/mnras/stu2446}, \href
  {https://ui.adsabs.harvard.edu/abs/2015MNRAS.447..927M} {447, 927}

\bibitem[\protect\citeauthoryear{{Milone} et~al.,}{{Milone}
  et~al.}{2015b}]{2015ApJ...808...51M}
{Milone} A.~P.,  et~al., 2015b, \mn@doi [\apj] {10.1088/0004-637X/808/1/51},
  \href {https://ui.adsabs.harvard.edu/abs/2015ApJ...808...51M} {808, 51}

\bibitem[\protect\citeauthoryear{{Milone} et~al.,}{{Milone}
  et~al.}{2017}]{2017MNRAS.464.3636M}
{Milone} A.~P.,  et~al., 2017, \mn@doi [\mnras] {10.1093/mnras/stw2531}, \href
  {https://ui.adsabs.harvard.edu/abs/2017MNRAS.464.3636M} {464, 3636}

\bibitem[\protect\citeauthoryear{{Mucciarelli}, {Origlia}  \&
  {Ferraro}}{{Mucciarelli} et~al.}{2007}]{2007AJ....134.1813M}
{Mucciarelli} A.,  {Origlia} L.,   {Ferraro} F.~R.,  2007, \mn@doi [\aj]
  {10.1086/522034}, \href
  {https://ui.adsabs.harvard.edu/abs/2007AJ....134.1813M} {134, 1813}

\bibitem[\protect\citeauthoryear{{Mucciarelli}, {Origlia}, {Ferraro}  \&
  {Pancino}}{{Mucciarelli} et~al.}{2009}]{2009ApJ...695L.134M}
{Mucciarelli} A.,  {Origlia} L.,  {Ferraro} F.~R.,   {Pancino} E.,  2009,
  \mn@doi [\apjl] {10.1088/0004-637X/695/2/L134}, \href
  {https://ui.adsabs.harvard.edu/abs/2009ApJ...695L.134M} {695, L134}

\bibitem[\protect\citeauthoryear{{Naiman}, {Ramirez-Ruiz}  \& {Lin}}{{Naiman}
  et~al.}{2018}]{2018MNRAS.478.2794N}
{Naiman} J.~P.,  {Ramirez-Ruiz} E.,   {Lin} D.~N.~C.,  2018, \mn@doi [\mnras]
  {10.1093/mnras/sty1198}, \href
  {https://ui.adsabs.harvard.edu/abs/2018MNRAS.478.2794N} {478, 2794}

\bibitem[\protect\citeauthoryear{{Nataf} et~al.,}{{Nataf}
  et~al.}{2019}]{2019AJ....158...14N}
{Nataf} D.~M.,  et~al., 2019, \mn@doi [\aj] {10.3847/1538-3881/ab1a27}, \href
  {https://ui.adsabs.harvard.edu/abs/2019AJ....158...14N} {158, 14}

\bibitem[\protect\citeauthoryear{{Niederhofer} et~al.,}{{Niederhofer}
  et~al.}{2017}]{2017MNRAS.465.4159N}
{Niederhofer} F.,  et~al., 2017, \mn@doi [\mnras] {10.1093/mnras/stw3084},
  \href {https://ui.adsabs.harvard.edu/abs/2017MNRAS.465.4159N} {465, 4159}

\bibitem[\protect\citeauthoryear{{Ostriker} \& {McKee}}{{Ostriker} \&
  {McKee}}{1988}]{astrowaves}
{Ostriker} J.~P.,  {McKee} C.~F.,  1988, \mn@doi [Reviews of Modern Physics]
  {10.1103/RevModPhys.60.1}, \href
  {http://adsabs.harvard.edu/abs/1988RvMP...60....1O} {60, 1}

\bibitem[\protect\citeauthoryear{{Palou{\v{s}}}, {W{\"u}nsch}  \&
  {Tenorio-Tagle}}{{Palou{\v{s}}} et~al.}{2014}]{2014ApJ...792..105P}
{Palou{\v{s}}} J.,  {W{\"u}nsch} R.,   {Tenorio-Tagle} G.,  2014, \mn@doi
  [\apj] {10.1088/0004-637X/792/2/105}, \href
  {https://ui.adsabs.harvard.edu/abs/2014ApJ...792..105P} {792, 105}

\bibitem[\protect\citeauthoryear{{Pancino} et~al.,}{{Pancino}
  et~al.}{2017}]{2017A&A...601A.112P}
{Pancino} E.,  et~al., 2017, \mn@doi [\aap] {10.1051/0004-6361/201730474},
  \href {https://ui.adsabs.harvard.edu/abs/2017A&A...601A.112P} {601, A112}

\bibitem[\protect\citeauthoryear{{Piotto} et~al.,}{{Piotto}
  et~al.}{2015}]{2015AJ....149...91P}
{Piotto} G.,  et~al., 2015, \mn@doi [\aj] {10.1088/0004-6256/149/3/91}, \href
  {https://ui.adsabs.harvard.edu/abs/2015AJ....149...91P} {149, 91}

\bibitem[\protect\citeauthoryear{{Rahner}, {Pellegrini}, {Glover}  \&
  {Klessen}}{{Rahner} et~al.}{2017}]{2017MNRAS.470.4453R}
{Rahner} D.,  {Pellegrini} E.~W.,  {Glover} S. C.~O.,   {Klessen} R.~S.,  2017,
  \mn@doi [\mnras] {10.1093/mnras/stx1532}, \href
  {https://ui.adsabs.harvard.edu/abs/2017MNRAS.470.4453R} {470, 4453}

\bibitem[\protect\citeauthoryear{{Raymond}, {Cox}  \& {Smith}}{{Raymond}
  et~al.}{1976}]{Raymond1976}
{Raymond} J.~C.,  {Cox} D.~P.,   {Smith} B.~W.,  1976, \mn@doi [\apj]
  {10.1086/154170}, \href {http://adsabs.harvard.edu/abs/1976ApJ...204..290R}
  {204, 290}

\bibitem[\protect\citeauthoryear{{Renzini}}{{Renzini}}{2008}]{2008MNRAS.391..354R}
{Renzini} A.,  2008, \mn@doi [\mnras] {10.1111/j.1365-2966.2008.13892.x}, \href
  {https://ui.adsabs.harvard.edu/abs/2008MNRAS.391..354R} {391, 354}

\bibitem[\protect\citeauthoryear{{Rosen}, {Lopez}, {Krumholz}  \&
  {Ramirez-Ruiz}}{{Rosen} et~al.}{2014}]{2014MNRAS.442.2701R}
{Rosen} A.~L.,  {Lopez} L.~A.,  {Krumholz} M.~R.,   {Ramirez-Ruiz} E.,  2014,
  \mn@doi [\mnras] {10.1093/mnras/stu1037}, \href
  {https://ui.adsabs.harvard.edu/abs/2014MNRAS.442.2701R} {442, 2701}

\bibitem[\protect\citeauthoryear{{Schiano}}{{Schiano}}{1985}]{1985ApJ...299...24S}
{Schiano} A.~V.~R.,  1985, \mn@doi [\apj] {10.1086/163680}, \href
  {https://ui.adsabs.harvard.edu/abs/1985ApJ...299...24S} {299, 24}

\bibitem[\protect\citeauthoryear{{Sedov}}{{Sedov}}{1959}]{1959sdmm.book.....S}
{Sedov} L.~I.,  1959, {Similarity and Dimensional Methods in Mechanics}

\bibitem[\protect\citeauthoryear{{Silich}}{{Silich}}{1992}]{silich1992}
{Silich} S.~A.,  1992, \mn@doi [\apss] {10.1007/BF00646764}, \href
  {http://adsabs.harvard.edu/abs/1992Ap%26SS.195..317S} {195, 317}

\bibitem[\protect\citeauthoryear{{Silich} \& {Tenorio-Tagle}}{{Silich} \&
  {Tenorio-Tagle}}{1998}]{1998MNRAS.299..249S}
{Silich} S.~A.,  {Tenorio-Tagle} G.,  1998, \mn@doi [\mnras]
  {10.1046/j.1365-8711.1998.01765.x}, \href
  {https://ui.adsabs.harvard.edu/abs/1998MNRAS.299..249S} {299, 249}

\bibitem[\protect\citeauthoryear{{Silich} \& {Tenorio-Tagle}}{{Silich} \&
  {Tenorio-Tagle}}{2017}]{Silich2017}
{Silich} S.,  {Tenorio-Tagle} G.,  2017, \mn@doi [\mnras]
  {10.1093/mnras/stw2879}, \href
  {https://ui.adsabs.harvard.edu/abs/2017MNRAS.465.1375S} {465, 1375}

\bibitem[\protect\citeauthoryear{{Silich} \& {Tenorio-Tagle}}{{Silich} \&
  {Tenorio-Tagle}}{2018}]{Silich2018}
{Silich} S.,  {Tenorio-Tagle} G.,  2018, \mn@doi [\mnras]
  {10.1093/mnras/sty1383}, \href
  {http://adsabs.harvard.edu/abs/2018MNRAS.478.5112S} {478, 5112}

\bibitem[\protect\citeauthoryear{{Silich}, {Tenorio-Tagle},
  {Mart{\'\i}nez-Gonz{\'a}lez}  \& {Turner}}{{Silich}
  et~al.}{2020}]{Silich2020}
{Silich} S.,  {Tenorio-Tagle} G.,  {Mart{\'\i}nez-Gonz{\'a}lez} S.,   {Turner}
  J.,  2020, \mn@doi [\mnras] {10.1093/mnras/staa705}, \href
  {https://ui.adsabs.harvard.edu/abs/2020MNRAS.494...97S} {494, 97}

\bibitem[\protect\citeauthoryear{{Sills}, {Dalessandro}, {Cadelano},
  {Alfaro-Cuello}  \& {Kruijssen}}{{Sills} et~al.}{2019}]{2019MNRAS.490L..67S}
{Sills} A.,  {Dalessandro} E.,  {Cadelano} M.,  {Alfaro-Cuello} M.,
  {Kruijssen} J.~M.~D.,  2019, \mn@doi [\mnras] {10.1093/mnrasl/slz149}, \href
  {https://ui.adsabs.harvard.edu/abs/2019MNRAS.490L..67S} {490, L67}

\bibitem[\protect\citeauthoryear{{Stevens} \& {Hartwell}}{{Stevens} \&
  {Hartwell}}{2003}]{2003MNRAS.339..280S}
{Stevens} I.~R.,  {Hartwell} J.~M.,  2003, \mn@doi [\mnras]
  {10.1046/j.1365-8711.2003.06184.x}, \href
  {https://ui.adsabs.harvard.edu/abs/2003MNRAS.339..280S} {339, 280}

\bibitem[\protect\citeauthoryear{{Tang} \& {Chevalier}}{{Tang} \&
  {Chevalier}}{2017}]{2017Tang}
{Tang} X.,  {Chevalier} R.~A.,  2017, \mn@doi [\mnras] {10.1093/mnras/stw2978},
  \href {http://adsabs.harvard.edu/abs/2017MNRAS.465.3793T} {465, 3793}

\bibitem[\protect\citeauthoryear{{Taylor}}{{Taylor}}{1946}]{1946RSPSA.186..273T}
{Taylor} G.~I.,  1946, \mn@doi [Proceedings of the Royal Society of London
  Series A] {10.1098/rspa.1946.0044}, \href
  {https://ui.adsabs.harvard.edu/abs/1946RSPSA.186..273T} {186, 273}

\bibitem[\protect\citeauthoryear{{Tenorio-Tagle} \&
  {Bodenheimer}}{{Tenorio-Tagle} \& {Bodenheimer}}{1988}]{1988ARA&A..26..145T}
{Tenorio-Tagle} G.,  {Bodenheimer} P.,  1988, \mn@doi [\araa]
  {10.1146/annurev.aa.26.090188.001045}, \href
  {https://ui.adsabs.harvard.edu/abs/1988ARA%26A..26..145T} {26, 145}

\bibitem[\protect\citeauthoryear{{Tenorio-Tagle}, {Franco}, {Bodenheimer}  \&
  {Rozyczka}}{{Tenorio-Tagle} et~al.}{1987a}]{1987A&A...179..219T}
{Tenorio-Tagle} G.,  {Franco} J.,  {Bodenheimer} P.,   {Rozyczka} M.,  1987a,
  \aap, \href {https://ui.adsabs.harvard.edu/abs/1987A%26A...179..219T} {179,
  219}

\bibitem[\protect\citeauthoryear{{Tenorio-Tagle}, {Bodenheimer}  \&
  {Rozyczka}}{{Tenorio-Tagle} et~al.}{1987b}]{1987A&A...182..120T}
{Tenorio-Tagle} G.,  {Bodenheimer} P.,   {Rozyczka} M.,  1987b, \aap, \href
  {https://ui.adsabs.harvard.edu/abs/1987A%26A...182..120T} {182, 120}

\bibitem[\protect\citeauthoryear{{Tenorio-Tagle}, {W{\"u}nsch}, {Silich}  \&
  {Palou{\v{s}}}}{{Tenorio-Tagle} et~al.}{2007}]{2007ApJ...658.1196T}
{Tenorio-Tagle} G.,  {W{\"u}nsch} R.,  {Silich} S.,   {Palou{\v{s}}} J.,  2007,
  \mn@doi [\apj] {10.1086/511671}, \href
  {https://ui.adsabs.harvard.edu/abs/2007ApJ...658.1196T} {658, 1196}

\bibitem[\protect\citeauthoryear{{Tenorio-Tagle}, {Mu{\~n}oz-Tu{\~n}{\'o}n},
  {Silich}  \& {Cassisi}}{{Tenorio-Tagle} et~al.}{2015}]{tenorio2015supernovae}
{Tenorio-Tagle} G.,  {Mu{\~n}oz-Tu{\~n}{\'o}n} C.,  {Silich} S.,   {Cassisi}
  S.,  2015, \mn@doi [\apjl] {10.1088/2041-8205/814/1/L8}, \href
  {http://adsabs.harvard.edu/abs/2015ApJ...814L...8T} {814, L8}

\bibitem[\protect\citeauthoryear{{Tenorio-Tagle}, {Silich}, {Palou{\v{s}}},
  {Mu{\~n}oz-Tu{\~n}{\'o}n}  \& {W{\"u}nsch}}{{Tenorio-Tagle}
  et~al.}{2019}]{2019ApJ...879...58T}
{Tenorio-Tagle} G.,  {Silich} S.,  {Palou{\v{s}}} J.,
  {Mu{\~n}oz-Tu{\~n}{\'o}n} C.,   {W{\"u}nsch} R.,  2019, \mn@doi [\apj]
  {10.3847/1538-4357/ab2455}, \href
  {https://ui.adsabs.harvard.edu/abs/2019ApJ...879...58T} {879, 58}

\bibitem[\protect\citeauthoryear{{Terlevich}, {Tenorio-Tagle}, {Franco}  \&
  {Melnick}}{{Terlevich} et~al.}{1992}]{terlevich1992starburst}
{Terlevich} R.,  {Tenorio-Tagle} G.,  {Franco} J.,   {Melnick} J.,  1992,
  \mn@doi [\mnras] {10.1093/mnras/255.4.713}, \href
  {http://adsabs.harvard.edu/abs/1992MNRAS.255..713T} {255, 713}

\bibitem[\protect\citeauthoryear{{Thornton}, {Gaudlitz}, {Janka}  \&
  {Steinmetz}}{{Thornton} et~al.}{1998}]{Thornton1998}
{Thornton} K.,  {Gaudlitz} M.,  {Janka} H.-T.,   {Steinmetz} M.,  1998, \mn@doi
  [\apj] {10.1086/305704}, \href
  {http://adsabs.harvard.edu/abs/1998ApJ...500...95T} {500, 95}

\bibitem[\protect\citeauthoryear{{Tomisaka} \& {Ikeuchi}}{{Tomisaka} \&
  {Ikeuchi}}{1986}]{1986PASJ...38..697T}
{Tomisaka} K.,  {Ikeuchi} S.,  1986, \pasj, \href
  {https://ui.adsabs.harvard.edu/abs/1986PASJ...38..697T} {38, 697}

\bibitem[\protect\citeauthoryear{{Truelove} \& {McKee}}{{Truelove} \&
  {McKee}}{1999}]{1999Mckee}
{Truelove} J.~K.,  {McKee} C.~F.,  1999, \mn@doi [\apj] {10.1086/313176}, \href
  {http://adsabs.harvard.edu/abs/1999ApJS..120..299T} {120, 299}

\bibitem[\protect\citeauthoryear{{Villanova}, {Geisler}, {Carraro}, {Moni
  Bidin}  \& {Mu{\~n}oz}}{{Villanova} et~al.}{2013}]{2013ApJ...778..186V}
{Villanova} S.,  {Geisler} D.,  {Carraro} G.,  {Moni Bidin} C.,   {Mu{\~n}oz}
  C.,  2013, \mn@doi [\apj] {10.1088/0004-637X/778/2/186}, \href
  {https://ui.adsabs.harvard.edu/abs/2013ApJ...778..186V} {778, 186}

\bibitem[\protect\citeauthoryear{{W{\"u}nsch}, {Silich}, {Palou{\v{s}}}  \&
  {Tenorio-Tagle}}{{W{\"u}nsch} et~al.}{2007}]{2007A&A...471..579W}
{W{\"u}nsch} R.,  {Silich} S.,  {Palou{\v{s}}} J.,   {Tenorio-Tagle} G.,  2007,
  \mn@doi [\aap] {10.1051/0004-6361:20077282}, \href
  {https://ui.adsabs.harvard.edu/abs/2007A&A...471..579W} {471, 579}

\bibitem[\protect\citeauthoryear{{W{\"u}nsch}, {Palou{\v{s}}}, {Tenorio-Tagle}
  \& {Ehlerov{\'a}}}{{W{\"u}nsch} et~al.}{2017}]{2017ApJ...835...60W}
{W{\"u}nsch} R.,  {Palou{\v{s}}} J.,  {Tenorio-Tagle} G.,   {Ehlerov{\'a}} S.,
  2017, \mn@doi [\apj] {10.3847/1538-4357/835/1/60}, \href
  {https://ui.adsabs.harvard.edu/abs/2017ApJ...835...60W} {835, 60}

\bibitem[\protect\citeauthoryear{{Yong} \& {Grundahl}}{{Yong} \&
  {Grundahl}}{2008}]{2008ApJ...672L..29Y}
{Yong} D.,  {Grundahl} F.,  2008, \mn@doi [\apjl] {10.1086/525850}, \href
  {https://ui.adsabs.harvard.edu/abs/2008ApJ...672L..29Y} {672, L29}

\bibitem[\protect\citeauthoryear{{Zhang}, {de Grijs}, {Li}  \& {Wu}}{{Zhang}
  et~al.}{2018}]{2018ApJ...853..186Z}
{Zhang} H.,  {de Grijs} R.,  {Li} C.,   {Wu} X.,  2018, \mn@doi [\apj]
  {10.3847/1538-4357/aaa428}, \href
  {https://ui.adsabs.harvard.edu/abs/2018ApJ...853..186Z} {853, 186}

\bibitem[\protect\citeauthoryear{{de Mink}, {Pols}, {Langer}  \& {Izzard}}{{de
  Mink} et~al.}{2009}]{2009A&A...507L...1D}
{de Mink} S.~E.,  {Pols} O.~R.,  {Langer} N.,   {Izzard} R.~G.,  2009, \mn@doi
  [\aap] {10.1051/0004-6361/200913205}, \href
  {https://ui.adsabs.harvard.edu/abs/2009A%26A...507L...1D} {507, L1}

\makeatother
\end{thebibliography}

%%%%%%%%%%%%%%%%%%%%%%%%%%%%%%%%%%%%%%%%%%%%%%%%%%

%%%%%%%%%%%%%%%%% APPENDICES %%%%%%%%%%%%%%%%%%%%%
\appendix
\section{The initial conditions for the SNR evolution}\label{Ap1}
We follow the evolution of the SNRs resulting from SN explosions occurring anywhere in the star-forming cloud from early times, when the reverse shock has not completely thermalized the ejected gas. As in \cite{2019MNRAS.488..978J}, the ejecta is included in our calculations assuming in all cases a linear profile for its velocity $v \propto r/t$, and a power-law distribution $\rho_{ej} \propto r^{-2}$ for the mass density. The ejecta mass is fixed at $M_{ej}=3$ M$_{\odot}$ and the explosion energy to $E_{0}=10^{51}$ erg. The initial conditions for the calculations are derived from the assumption that a fraction $\alpha$ of the initial explosion energy $E_{0}$ has already transformed into kinetic and thermal energies of the shocked gas (i.e., $\alpha=1$ implies complete thermalization, while $\alpha=0$ means that the reverse shock is not yet moving towards the center of the explosion).  Here, all calculations were provided for the case $\alpha=0.3$. This assumption allows one (as shown by \citealt{2019MNRAS.488..978J}) to estimate the initial time $t_{0}$, the energies $E_{\textrm{k,sw}}\left(t_{0}\right)$, $E_{\textrm{k,ej}}\left(t_{0}\right)$, $E_{\textrm{k,free}}\left(t_{0}\right)$, $E_{\textrm{th}}\left(t_{0}\right)$ as well as the starting positions and velocities of the leading and reverse shocks.  These are the initial conditions required to solve equations (\ref{fun1}-\ref{thermalE}) for each Lagrangian element. 
\section{The central pressure in a gaussian cloud}\label{Ap2}
The turbulent pressure is given by the equation (\ref{eq2}) and the gas density by equation (\ref{eq1}). The total mass within radius $r$ is determined by:
\begin{equation}\label{Aeq3}
M\left(r \right)= M_{\textrm{tot}} \left[ \textrm{erf}\left(\frac{r}{\sqrt{2}R_{c}} \right)-\left(\frac{2}{\pi} \right)^{1/2}\frac{r}{R_{c}}\textrm{exp} \left( -\frac{1}{2}\frac{r^{2}}{R_{c}^{2}}\right) \right].
\end{equation}
Then, using equations (\ref{eq1})-(\ref{Aeq3}) in equation (\ref{eq2}) and integrating:
\begin{equation}
\begin{split}
P_{g}\left(r \right) & = P_{\textrm{amb}}+ \frac{-G M_{\textrm{gas}}^{2}}{\left(1-\epsilon_{\textrm{1G}}\right)\left( 2 \pi \right)^{\frac{3}{2}}R_{c}^{4}} \times \\
& \int_{\frac{Rsc}{R_{c}}}^{\frac{r}{R_{c}}} \left[ \textrm{erf} \left(\frac{s}{\sqrt{2}} \right) - \sqrt{\frac{2}{\pi}}s \textrm{exp} \left(\frac{-s^{2}}{2}\right) \right] \textrm{exp} \left(  \frac{-s^{2}}{2}\right)\frac{ds}{s^2},
\end{split}
\end{equation}
where $R_{sc}$ is the outer boundary of the cloud and $P_{\textrm{amb}}=P_{g}\left(R_{\textrm{SC}} \right)$ is the thermal pressure of the ambient gas outside the gas cloud. The central pressure is then:
\begin{equation}\label{centralPressure}
P_{0}=\beta \frac{G M_{\textrm{gas}}^{2}}{\left(1-\epsilon_{\textrm{1G}}\right)\left( 2 \pi \right)^{\frac{3}{2}}R_{c}^{4}}+P_{\textrm{amb}},
\end{equation}
with:
\begin{equation}
\beta=\int_{0}^{\frac{Rsc}{R_{c}}} \left[ \textrm{erf} \left(\frac{s}{\sqrt{2}} \right) -\sqrt{\frac{2}{\pi}}s \textrm{exp} \left(\frac{-s^{2}}{2}\right)\right]\textrm{exp} \left(  \frac{-s^{2}}{2}\right)\frac{ds}{s^2}.
\end{equation}
As one can note, $\beta$ depends on the quotient $R_{sc}/R_{c}$, however $R_{sc}/R_{c}>>1$ and therefore $\beta$ can be approximated as:
\begin{equation}
\begin{split}
\beta= & \int_{0}^{\infty} \left[ \textrm{erf} \left(\frac{s}{\sqrt{2}} \right) -\sqrt{\frac{2}{\pi}}s \textrm{exp} \left(\frac{-s^{2}}{2}\right)\right] \times \\
& \textrm{exp} \left(  \frac{-s^{2}}{2}\right)\frac{ds}{s^2}\approx 0.171227492145 \approx 0.17.
\end{split}
\end{equation}
Therefore, from equation (\ref{centralPressure}):
\begin{equation}\label{turbP1}
P_{0}=0.17 \frac{G M_{\textrm{gas}}^2}{\left( 2 \pi \right)^{3/2}\left(1-\epsilon_{\textrm{1G}} \right)R_{c}^{4}}+P_{\textrm{amb}}.
\end{equation}
The second term on equation (\ref{turbP1}) is orders of magnitude smaller than the first term for the initial conditions used here (see section \ref{clusterModel}), hence, the central pressure is approximated as: 
\begin{equation}\label{turbPfinal}
P_{0} \approx 0.17 \frac{G M_{\textrm{gas}}^2}{\left( 2 \pi \right)^{3/2}\left(1-\epsilon_{\textrm{1G}} \right)R_{c}^{4}}.
\end{equation}

\section{Multiple supernovae within massive and compact star clusters}\label{Ap3}
Unlike stellar winds, supernovae are discrete events. Let us assume a constant type II SN rate. Then, one can obtain the time span between consecutive SN explosions $\Delta t_{\textrm{SN}}$ within the blowout radius $R_{\textrm{blow}}$. Assuming that feedback from supernovae last $40 \textrm{Myr}$:
\begin{equation}
\Delta t_{\textrm{SN}}=\frac{3 \times 10 ^{4} \textrm{ yr}}{\left(M_{\textrm{1G}}/10^{5} M_{\odot} \right)\textrm{f}_{\textrm{m}}},
\end{equation}
where:
\begin{equation}
\textrm{f}_{\textrm{m}}=\textrm{erf}\left(\frac{R_{\textrm{blow}}}{\sqrt{2}R_{c}} \right)-\sqrt{\frac{2}{\pi}}\frac{R_{\textrm{blow}}}{Rc}\textrm{exp} \left[-\frac{1}{2}\left(\frac{R_{\textrm{blow}}}{Rc} \right)^{2} \right].
\end{equation}  
 
The SN cooling time $t_{\textrm{cool,SN}}$ is inversely proportional to the gas density \citep[e.g.][]{Thornton1998, LiOstriker,KimOstriker}. Fig. \ref{ap_fig1} presents $t_{\textrm{cool,SN}}$ as a function of the gas density for several numerical calculations. The solid line in Fig. \ref{ap_fig1} is the best fit to the data given by the equation:
\begin{equation}\label{ap3_eq1}
t_{\textrm{cool,SN}}=2.4 \times 10^{3} n_{0,5}^{-0.49} \textrm{ yr},
\end{equation}
where $n_{0,5}=n_{0}/(10^{5} \textrm{ cm}^{-3})$ is the ambient gas number density, which depends on the 1G mass $M_{\textrm{1G}}$, star formation efficiency $\epsilon_{\textrm{1G}}$ and on the star cluster core radius $R_{\textrm{c}}$ (see equation \ref{eq1}):
\begin{equation}\label{ap3_eq2}
n_{0}=\frac{\left(1- \epsilon_{\textrm{1G}}\right)M_{\textrm{1G}}}{\left(2 \pi \right)^{3/2}\mu \epsilon_{\textrm{1G}}R_{c}^{3}}.
\end{equation} 
\begin{figure}
	\includegraphics[width=\columnwidth]{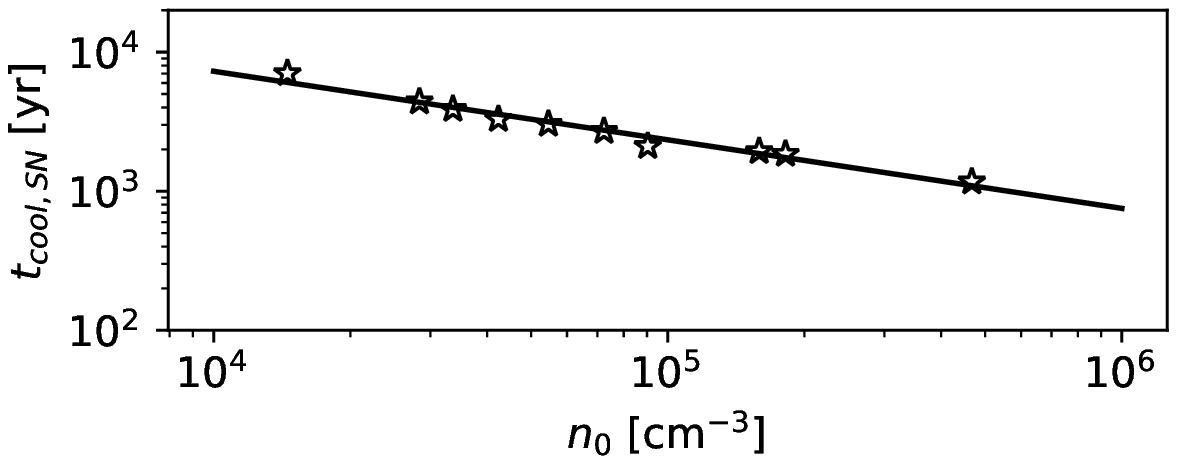}
    \caption{The supernova cooling time $t_{\textrm{cool,SN}}$ as a function of the gas density $n_{0}$. Numerical results are indicated by the star symbols while the best power-law fit is shown by the solid line. }
    \label{ap_fig1}
\end{figure}
\begin{figure}
	\includegraphics[width=\columnwidth]{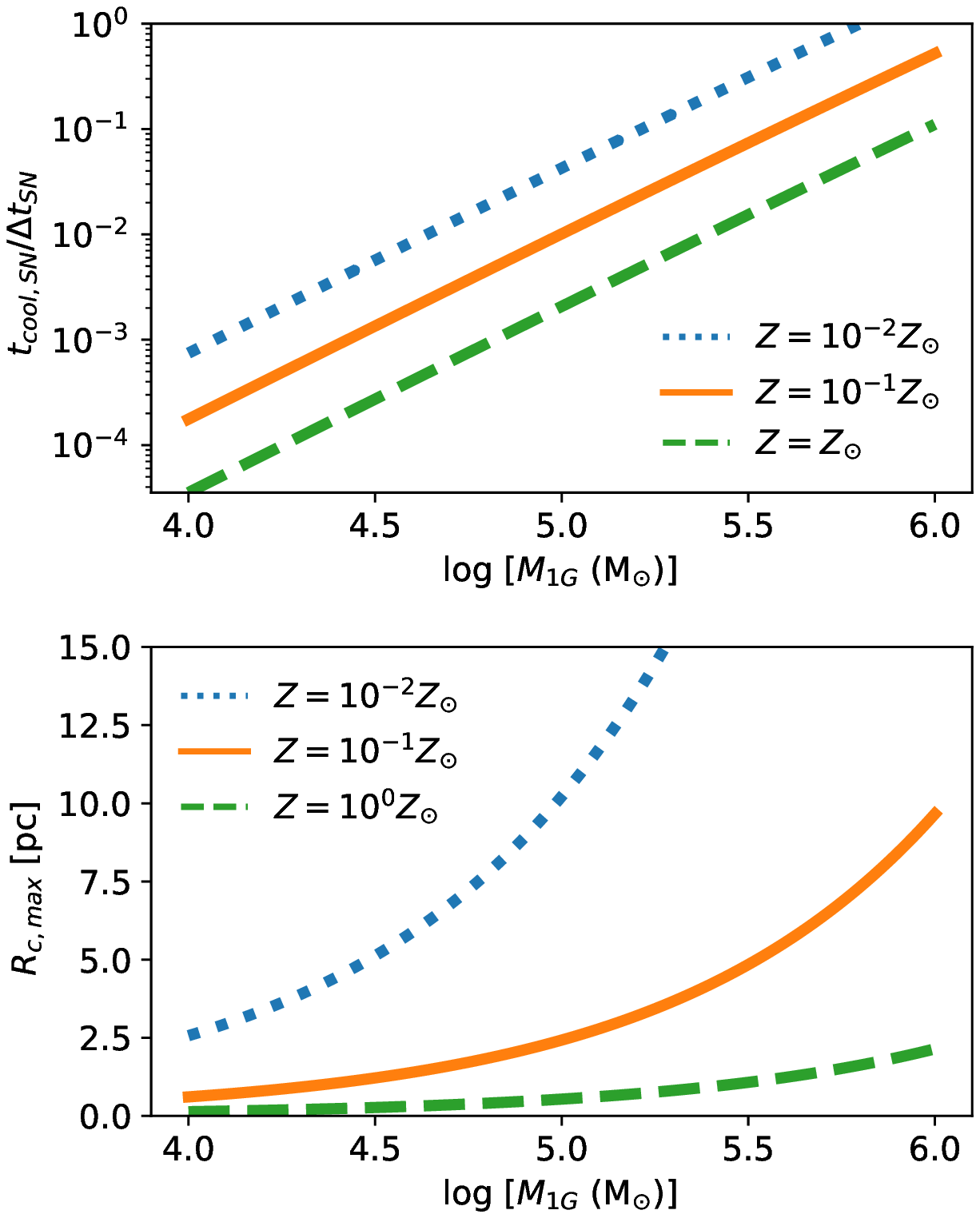}
    \caption{Top: the SN cooling time $t_{\textrm{cool,SN}}$ to time span $\Delta t_{\textrm{SN}}$ between consecutive SN ratio as a function of the 1G stellar mass $M_{\textrm{1G}}$, for different cloud metallicities (inset legend). The ratio $t_{\textrm{cool,SN}}/\Delta t_{\textrm{SN}}$ is calculated for the maximum allowed star formation efficiency and core radius for each case.  Bottom: The maximum allowed core radius $R_{\textrm{c,max}}$ as a function of the 1G stellar mass. The different curves correspond to the sames metallicity cases as the top panel. }
    \label{ap_fig2}
\end{figure}

The $t_{\textrm{cool,SN}}$ to $\Delta t_{\textrm{SN}}$ ratio calculated for different gas metallicities and 1G masses is presented on the top panel of Fig. \ref{ap_fig2}. Here different curves were calculated for the maximum allowed star formation efficiencies and core radii $R_{\textrm{c,max}}$ ($\epsilon_{\textrm{1G,max}}$ =0.027, 0.11 and 0.42 for Z$=10^{-2}$  Z$_{\odot}$, $10^{-1}$  Z$_{\odot}$ and Z$_{\odot}$, respectively, see section \ref{enrichment_section}, $\beta=1$ in equation \ref{WIND_RADIUS2}):
\begin{equation}\label{ap3_eq3}
R_{\textrm{c,max}}=\left[ \frac{M_{\textrm{gas,6}}^{3}}{\lambda^{3} \epsilon_{\textrm{1G}}\left( 1-\epsilon_{\textrm{1G}}\right)}\right]^{1/5}.
\end{equation}
The bottom panel on Fig. \ref{ap_fig2} presents the corresponding $R_{\textrm{c,max}}$. The $t_{\textrm{cool,SN}}/\Delta t_{\textrm{SN}}$ ratios shown in the top panel of Fig. \ref{ap_fig2} are upper limits for each 1G stellar mass because smaller efficiencies and core radii lead to larger gas densities and consequently, to smaller cooling times $t_{\textrm{cool,SN}}$ (equations \ref{ap3_eq1} and \ref{ap3_eq2}). Note that $t_{\textrm{cool,SN}}/\Delta t_{\textrm{SN}}<1$ for reasonable values of the core radius as those used in all our simulations (see Figs. \ref{Fe_increase} and \ref{Fe_increase_b}). Therefore SNRs overlapping is very unlikely.

% Don't change these lines

\bsp	% typesetting comment
\label{lastpage}
\end{document}